\documentclass[aps,prd,twocolumn,nofootinbib,showpacs,groupedaddress,%superscriptaddress,
preprintnumbers,amsmath,amssymb,floatfix]{revtex4-2}

\usepackage[bookmarks,linktocpage,colorlinks = true,linkcolor = blue,urlcolor  = blue,citecolor = blue,anchorcolor = green,hyperindex = true,hyperfigures]{hyperref}

\usepackage{graphicx} 
\usepackage{dcolumn}   
\usepackage{bm}        
\usepackage{amssymb}   
\usepackage{multirow}
\usepackage{feynmf}
\usepackage{slashed}
\usepackage{natbib}
\usepackage{epstopdf}
\usepackage[usenames]{color}
\usepackage{float}

\newcommand{\be}{\begin{equation}}
\newcommand{\ee}{\end{equation}}
\newcommand{\ba}{\begin{eqnarray}}
\newcommand{\ea}{\end{eqnarray}}

\newcommand{\nB}{n_{\mathrm{B}}}
\newcommand{\nsat}{n_{\mathrm{sat}}}

\newcommand{\muB}{\mu_{\mathrm{B}}}
\newcommand{\muI}{\mu_{\mathrm{I}}}
\newcommand{\muH}{\mu_{\mathrm{H}}}
\newcommand{\muL}{\mu_{\mathrm{sat}}}
\newcommand{\muq}{\mu_{\mathrm{q}}}

\newcommand{\musat}{\mu_{\mathrm{sat}}}

\newcommand{\nL}{n_{\mathrm{sat}}}
\newcommand{\nH}{n_{\mathrm{H}}}

\newcommand{\ChiEFT}{$\chi$EFT}

\newcommand{\Nc}{N_{\mathrm{c}}}
\newcommand{\Nf}{N_{\mathrm{f}}}
\newcommand{\ZB}{Z_{\mathrm{B}}}
\newcommand{\ZI}{Z_{\mathrm{I}}}
\newcommand{\PB}{P_{\mathrm{B}}}
\newcommand{\PI}{P_{\mathrm{I}}}
\newcommand{\PL}{P_{\mathrm{sat}}}
\newcommand{\PH}{P_{\mathrm{H}}}

\newcommand{\calD}{\mathcal{D}}

\newcommand{\calO}{\mathcal{O}}
\renewcommand{\Re}{\mathop{\mathrm{Re}}}

\begin{document}

\preprint{INT-PUB-23-043}
\title{Bounds on the Equation of State from QCD Inequalities and Lattice QCD}
\author{Yuki~Fujimoto}
\email{yfuji@uw.edu}
\affiliation{Institute for Nuclear Theory, University of Washington, Box 351550, Seattle, WA, 98195, USA}
\author{Sanjay~Reddy}
\affiliation{Institute for Nuclear Theory, University of Washington, Box 351550, Seattle, WA, 98195, USA}

\date{\today}

\begin{abstract}
We derive robust bounds on the equation of state (EoS) at finite baryon chemical potential using QCD inequalities and input from recent lattice-QCD calculations of thermodynamic properties of matter at nonzero isospin chemical potential.
We use lattice data to deduce an upper bound on the baryon density of the symmetric nuclear matter at a given baryon chemical potential and a lower bound on the pressure as a function of the energy density. We also use constraints from perturbative calculations of the QCD EoS at high density derived in earlier work and causality to delineate robust bounds on the EoS of isospin symmetric matter at densities relevant to heavy-ion collisions.
\end{abstract}

\maketitle

\section{Introduction}

Recent studies have provided useful constraints on the equation of state (EoS) of dense matter using input from astrophysics and nuclear physics.
Several authors have shown that it is possible to combine measurements of heavy neutron star masses, neutron star radii, and tidal deformability to constrain the pressure of neutron-rich matter at baryon density in the range 2-4 $\nsat$, where $\nsat\simeq 0.16~\mathrm{fm}^{-3}$ is the saturation density inside nuclei.
At lower density, where nuclear matter is non-relativistic and dilute, nuclear Hamiltonians derived using phenomenological considerations and Chiral Effective Field Theory (\ChiEFT) now provide useful constraints on the EoS, and a comprehensive discussion of these calculations and results are reviewed in \cite{Drischler:2021kxf}.
At much higher baryon density, for $\nB \gtrsim 40~\nsat$, the typical momentum scale for quark and gluon interactions become much larger than $\Lambda_{\mathrm{QCD}} \simeq 200 $ MeV and perturbative QCD (pQCD) calculations provide reliable and stringent constraints on the EoS as reviewed in \cite{Ghiglieri:2020dpq}.
Further, in Ref.~\cite{Komoltsev:2021jzg}, it was shown that thermodynamic consistency and stability conditions could be used to extrapolate the pQCD constraints to lower density.

The tightest constraints on the EoS of neutron-rich matter, which is characterized by a large isospin asymmetry due to constraints imposed by charge neutrality and beta-equilibrium, are obtained from observations of neutron star structure.
There is a one-to-one correspondence between the mass-radius relationship of neutron stars and the EoS through Einstein equation~\cite{1992ApJ...398..569L} (and similarly for other observables such as tidal deformability), and hence astrophysical measurements provide robust bounds on the EoS. Indeed, it has been shown that the large portion of the allowed region of the EoS is strictly ruled out by the tidal deformability bound from the GW170817 event and the existence of the two-solar-mass pulsars~\cite{Kurkela:2014vha, Annala:2017llu, Landry:2020vaw, Annala:2021gom}.

As for isospin symmetric matter, experimental measurements of the isoscalar giant monopole resonances in nuclei provide strong constraints on the incompressibility coefficient of symmetric nuclear matter EoS at saturation density (see, e.g., Ref.~\cite{Piekarewicz:2009gb} for a review), but reliable constraints at higher density have been elusive.
Although there has been progress in identifying several EoS-sensitive observables in heavy-ion collisions that access high baryon density, an interpretation of the data has been difficult.  The systematic uncertainties associated with the hadronic transport models needed in this context remain poorly understood, and the EoS constraints derived using them (see, e.g.~\cite{Danielewicz:2002pu, Oliinychenko:2022uvy, Sorensen:2023zkk}) are not as robust as the astrophysical constraints on isospin asymmetric matter.

The purpose of this study is to demonstrate that we can use lattice-QCD calculations of thermodynamic properties at $\muI>0$ to derive useful and robust bounds on the EoS of isospin symmetric matter at $\muB >0$ and low temperature ($\muI$ and $\muB$ are isospin and baryon chemical potentials, respectively). Intriguingly, although the ground state of the matter at $\muB=0$ and nonzero $\muI$, which is characterized by a Bose condensate of pions for $\muI >  m_\pi$ ($m_\pi$ is the pion mass)~\cite{Son:2000xc,*Son:2000by}, is very different from baryonic matter at nonzero $\muB$, a QCD inequality that relates the pressures of matter at nonzero $\muI$ and $\muB$ derived by Cohen in Refs.~\cite{Cohen:2003ut, *Cohen:2004qp}, allows us to derive this bound. We employ results from recent lattice-QCD calculations at $\muI >0$~\cite{Abbott:2023coj} to obtain an upper bound on the pressure as a function of $\muB$.

The QCD inequalities, pioneered by the seminal works dating back to four decades ago~\cite{Weingarten:1983uj, Witten:1983ut, Vafa:1983tf}, relate different correlation functions without explicitly evaluating them. They are derived from inequalities among the integrands in the path integral expressions;  using the fact that the path integral measure is positive, path-integrated quantities also satisfy inequality relations (see also Ref.~\cite{Nussinov:1983vh} for a Hamiltonian variation approach). The QCD inequalities have been successful in discussing the symmetry-breaking patterns, comparing hadron masses, etc.\ in the \emph{vacuum} (see Ref.~\cite{Nussinov:1999sx} for a comprehensive review).
In contrast, at nonzero chemical potential, the QCD inequalities generally cannot hold between path-integrated quantities because the Fermion determinant becomes complex-valued, and the path integral measure is not positive -- and is widely known as the Fermion sign problem.
However, there is an exceptional case where one can still obtain the positive path integral measure with nonzero chemical potential;
it is QCD at nonzero $\muI$, which can be regarded as a complex phase-quenched theory for QCD at nonzero $\muB$~\cite{Alford:1998sd, Son:2000xc,*Son:2000by} (see also \cite{Moore:2023glb}).
From this fact, one can put an upper bound on the path integral of QCD at nonzero $\muB$ from that of QCD at nonzero $\muI$ (see, e.g., Ref.~\cite{Hidaka:2011jj} for an application of the QCD inequality at nonzero $\muI$).

The positivity of the path integral measure in QCD at nonzero $\muI$ circumvents the sign problem and there have been several lattice studies of the phase structure and thermodynamic properties of the two-flavor isospin matter at nonzero $\muI$ ~\cite{Kogut:2002tm, Kogut:2002zg, Kogut:2004zg, deForcrand:2007uz, Detmold:2008fn, Cea:2012ev, Detmold:2012wc, Detmold:2012pi, Endrodi:2014lja, Brandt:2017oyy, Brandt:2018omg, Brandt:2022hwy, Brandt:2023kev, Abbott:2023coj}. Apart from the QCD-like theory with $\Nc = 2$  (see, e.g., \cite{Hands:2006ve, Cotter:2012mb, Boz:2019enj, Begun:2022bxj, Iida:2022hyy}), QCD with $\Nc = 3$, $\muB=0$ and $\muI>0$ is the only system for which lattice calculation of the EoS at nonzero chemical potential around vanishing temperature is feasible. A recent lattice QCD  calculation was able to construct states with a large number of pions (6144) corresponding to $\muI >0$ and negligible temperature and measure their thermodynamic properties~\cite{Abbott:2023coj}. We use their results to constrain the thermodynamic properties of matter with $\muB>0$ at vanishing temperature. To our knowledge, this is the first example of a lattice bound on the EoS at nonzero $\muB$ and $T\approx 0$.

The bound we derive applies to isospin symmetric matter at nonzero $\muB$ with zero net strangeness. Such matter is interesting because it is realized in heavy-ion collisions and is relevant to the QCD critical point searches in heavy-ion collisions. 
The critical point is the endpoint of a conjectured first-order line in the $\muB - T$ plane of isospin symmetric matter (for a review, see \cite{Stephanov:2004wx, Bzdak:2019pkr}). If the first-order phase transition persists at low temperatures, the baryon density would be discontinuous across it.   
To constrain the jump in density, we translate constraints on the pressure to a baryon number density constraint using the integral constraint method developed in Ref.~\cite{Komoltsev:2021jzg}.
This method also allows us to derive constraints on the pressure as a function of the baryon energy density at nonzero $\muB$ and low temperature. The bounds we find may not seem stringent but robustly exclude a soft EoS characterized by a low sound speed $v_s^2 \lesssim 0.2$ for $\muB$ in the range  $1$-$2$ GeV. We find that they are competitive with bounds derived from robust extrapolations of pQCD that rely on thermodynamic consistency, stability, and causality conditions~\cite{Komoltsev:2021jzg}. The interplay between these independent bounds could provide guidance for both Lattice-QCD and pQCD.

The paper is organized as follows.
In Sec.~\ref{sec:ineq}, we review the inequalities that compare the QCD thermodynamics at nonzero $\muB$ and $\muI$ in detail. We show how recent lattice results constrain the EoS in the $\muB - P$ plane in Section~\ref{sec:Pmu}. In Sec.~\ref{sec:mun}, we use constraints on the $\muB - P$ plane to constrain  $\nB(\muB)$ and eventually $P(\varepsilon)$.
In doing so, we express the pressure as an integral of the baryon density and specify the constant of integration using empirical information about nuclear matter at the saturation point to obtain a lower bound on the pressure as a function of the energy density. 
In Sec.~\ref{sec:pQCD}, we use pQCD constraints on the high-density EoS to specify the constant of integration and isospin-QCD lattice data to obtain an upper bound on the pressure as a function of the energy density. In section~\ref{sec:integ}, we compare results obtained in the previous sections with the pQCD integral constraint derived earlier from the thermodynamic consistency, stability, and causality conditions~\cite{Komoltsev:2021jzg}.

\section{QCD inequalities at nonzero chemical potentials}
\label{sec:ineq}

Here, we review QCD inequalities at nonzero $\muB$ and $\muI$ and derive the relationship between QCD partition functions of the baryonic matter and the isospin matter, which are denoted as $\ZB(\muB)$ and $\ZI(\muI)$, respectively. The derivation is based on a Euclidean path integral representation and was presented in Refs.~\cite{Cohen:2003ut, *Cohen:2004qp} (see also Ref.~\cite{Cohen:2003kd}).
In the following, we consider QCD in an Euclidean space with $\Nf=2$ degenerate flavors.
We will specify the temperature to be zero, but the inequalities shown below also hold at any temperature.

\subsection{Partition function with nonzero baryon chemical potential}
The Dirac operator, $\calD(\muq)$, at a nonzero real-valued quark chemical potential, $\muq$, is given by 
\begin{equation}
    \label{eq:Diracop}
    \calD(\mu) \equiv \slashed{D} + m - \muq \gamma_0\,,
\end{equation}
where the covariant derivative, $\slashed{D} \equiv \slashed{\partial} + ig\slashed{A}$, is a skew-Hermitian operator, i.e.\ $\slashed{D}^\dagger = - \slashed{D}$.
Furthermore, due to the skew-Hermiticity of $\slashed{D}$, the Dirac operator at $\muq=0$ becomes pseudo-Hermitian by $\gamma_5$, and 
\begin{equation}
    \gamma_5 \calD(\muq=0) \gamma_5 = -\slashed{D} + m = \calD^\dagger(\muq=0)\,.    
\end{equation}
When $\muq=0$, this $\gamma_5$-pseudo-Hermiticity guarantees a positive path integral measure with $\det \calD(\muq=0)\geq 0$;  this positivity is key to deriving QCD inequalities for path integrated quantities. In contrast, at $\muq \neq 0$, the $\gamma_5$-pseudo-Hermiticity is lost because 
\begin{equation}
    \label{eq:phlost}
    \gamma_5 \calD(\muq) \gamma_5 = \calD^\dagger(-\muq) \neq \calD^\dagger(\muq)\,.
\end{equation}
 and consequently, the path integral measure is no longer positive.

For QCD with two flavors, the partition function $\ZB$ in the path integral representation is given by
\begin{align}
  \label{eq:ZB}
  \ZB(\muB) = \int [dA]
  \left[\det\calD(\tfrac{\muB}{\Nc})\right]^{2} e^{-S_{\rm G}}\,,
\end{align}
where $S_{\rm G}$ is the Euclidean action of QCD in the gauge sector.
In general, the fermion determinant in the above expression is complex. 
However, using  the charge conjugation symmetry that requires $\ZB(\muB) = \ZB(-\muB)$ and the following relation 
derived from Eq.~\eqref{eq:phlost}
\begin{equation}
    \label{eq:detast}
    \det\calD(-\muq) = \det \gamma_5 \calD(-\muq) \gamma_5 = \left[\det \calD(\muq)\right]^\ast\,.
\end{equation}
it can be shown that 
\begin{align}
  \label{eq:ZBreal}
  \ZB(\muB) = \int [dA] \Re \left[\det\calD(\tfrac{\muB}{\Nc})\right]^2 e^{-S_{\rm G}}\,.
\end{align}
As should be expected on physical grounds since the partition function should be real-valued function  \cite{Cohen:2003ut, *Cohen:2004qp} (see also Refs.~\cite{deForcrand:2002pa, deForcrand:2009zkb, Hsu:2010zza, Giordano:2020roi, Borsanyi:2021hgr})

\subsection{Partition function with nonzero isospin chemical potential}

The path integral representation of the partition function of $u$ and $d$ quarks at finite $\muI$ and $\muB=0$ is given by 
\begin{align}
  \label{eq:ZI}
  \ZI(\muI) &= \int [dA]
  \det\calD(\tfrac{\muI}{2}) 
  \det\calD(-\tfrac{\muI}{2}) e^{-S_{\rm G}}\,,
\end{align}
where $\calD(\muq)$ is the Dirac operator defined in Eq.~\eqref{eq:Diracop}.
The arguments of the fermion determinants have opposite signs $\pm\muI$ as $u$ and $d$ quarks have opposite (third components of) isospins $I_3$.
From the relation~\eqref{eq:detast}, $\ZI(\muI)$ can be rewritten as
\begin{align}
    \label{eq:ZIabs}
    \ZI(\muI) = \int [d A]
  \left|\det\calD(\tfrac{\muI}{2})\right|^2  e^{-S_{\rm G}} \,.
\end{align}
The positivity of the path integral at finite $\muI$ measure mentioned earlier is now explicit in Eq.~\eqref{eq:ZIabs}. We note that QCD at nonzero $\muI$ can also be regarded as the \emph{phase-quenched} theory of two-flavor QCD at nonzero $\muB$ in which the complex phase of the fermion determinant is discarded. This is quite distinct from the quenched approximation in which the entire fermion determinant is neglected.

\subsection{QCD inequalities}
From the relation $\Re z^2 \leq |z^2| = |z|^2$, the following inequality holds
\begin{equation}
    \Re \left[\det\calD(\tfrac{\muB}{\Nc})\right]^2 \leq \left|\det\calD(\tfrac{\muB}{\Nc})\right|^2 \,.
\end{equation}
From this inequality, we get an upper bound on $\ZB(\muB)$
\begin{align}
  \label{eq:ZBleq}
 \ZB(\muB) \leq \int [dA] \left|\det\calD(\tfrac{\muB}{\Nc})\right|^2 e^{-S_{\rm G}}\,.
\end{align}
The LHS and RHS differ by the phase of the determinant, so the inequality is saturated when the phase is unity.
The RHS can be recast as $\ZI(\muI)$ by mapping $\muB$ to $\muI$ with an appropriate prefactor, which is $\muI = 2 \muB / \Nc$.
We see that Eq.~\eqref{eq:ZBleq} combined with Eq.~\eqref{eq:ZIabs} yields a useful inequality
\begin{align}
    \label{eq:cohen_Z}
  \ZB(\muB) \leq \ZI\left(\muI = \frac{2 \muB}{\Nc}\right)\,, 
\end{align}
which was first derived by Cohen in Ref.~\cite{Cohen:2003ut, *Cohen:2004qp}.

By taking the logarithm of this inequality, one obtains an upper bound on the pressure of the baryonic matter at a given $\muB$ in terms of the pressure of isospin matter $\muI (= 2 \muB/\Nc)$
\begin{align}
  \label{eq:cohen}
  \PB(\muB) \leq \PI\left(\muI = \frac{2\muB}{\Nc}\right)\,. 
\end{align}
This inequality will eventually be saturated at asymptotically high density, as can be seen in the perturbative expressions of the pressure at nonzero $\muB$ and $\muI$ as they are identical up to order $\alpha_s^2$, where $\alpha_s$ is the strong coupling constant.
The difference appears at $\mathcal{O}(\alpha_s^3)$~\cite{Moore:2023glb}.

\subsection{An inequality for baryonic matter with isospin imbalance}
The pressure inequality derived in the preceding discussion applies to isospin-symmetric baryonic matter.
From the convexity condition of the pressure derived in Ref.~\cite{Lee:2004hc} and given by 
\begin{equation}
    P(\muB, 0) \leq P(\muB, \muI) \leq \tfrac12 \left[P(\muB + \muI, 0) + P(\muB - \muI, 0)\right]\,.
\end{equation}
 one can derive a bound on the pressure at nonzero $\muB$ and finite isospin imbalance $\muI$, denoted as $P(\muB, \muI)$. Given the relations between baryonic and isospin pressures and the general pressure with an arbitrary isospin imbalance, $\PB(\muB) = P(\muB, 0)$ and $\PI(\muI) = P(0, \muI)$, and by combining with the QCD inequalities above, we obtain
\begin{equation}
    P(\muB, \muI)
    \leq \tfrac12 \left[ P(0, \tfrac{2}{\Nc}(\muB + \muI)) + P(0, \tfrac{2}{\Nc}(\muB - \muI))\right]\,.
\end{equation}
This can, in principle, be applied to the neutron star matter where the charge neutrality and the beta-equilibrium condition are fulfilled with nonzero $\muI$.
In practice, however, this inequality requires the value of $\muI$ as a function of $\muB$, which we cannot know from the current neutron star observations unless we assume some model.

\section{Lattice-QCD bound on the baryonic matter pressure}
\label{sec:Pmu}

From the inequality~\eqref{eq:cohen}, the lattice-QCD calculation of the isospin matter EoS puts an upper bound for the two-flavor symmetric matter EoS.

\begin{figure}[ht]
    \centering
    \includegraphics[width=0.95\columnwidth]{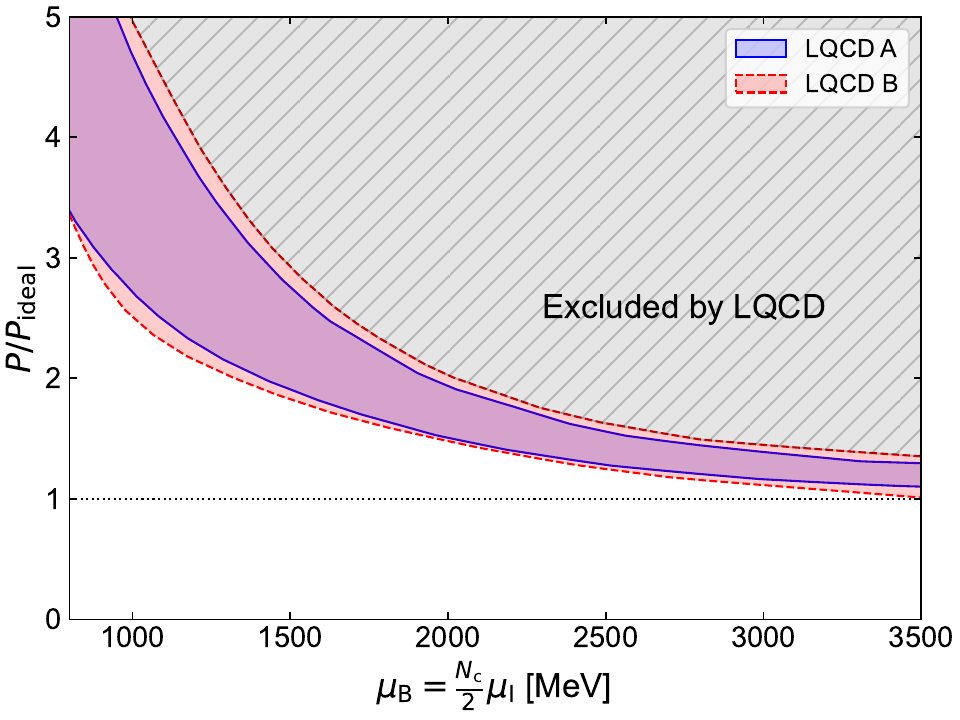}
    \caption{Pressure of the isospin matter.  The pressure is normalized by the ideal quark gas value $P_{\rm ideal} \equiv \Nc \Nf \muq^4/(12 \pi^2)$ with $\muq \equiv \muB / \Nc = \muI / 2$.  The grey hatched region is excluded by the isospin lattice-QCD data.}
    \label{fig:ppid}
\end{figure}

In Fig.~\ref{fig:ppid}, we plot the lattice-QCD results of the isospin matter pressure from Ref.~\cite{Abbott:2023coj}.
The blue and red shaded regions marked with LQCD A and LQCD B in Fig.~\ref{fig:ppid} are the results sampled from different ensembles at nearly vanishing temperature, $T \sim 23~\text{MeV}$ and $17~\text{MeV}$ for the ensembles A and B, respectively.
The $x$-axis is rescaled as $\muI \to \muB = (\Nc/2) \muI$.
The normalized pressure $P/P_{\rm ideal}$ is read out from the lattice data in Ref.~\cite{Abbott:2023coj} by multiplying $(1/3 - \Delta)$ and $3\varepsilon/\varepsilon_{\rm ideal}$, where $\Delta \equiv 1/3 - P/\varepsilon$ and $3 P_{\rm ideal} = \varepsilon_{\rm ideal}$.
The pressure of the ideal quark gas is given by $P_{\rm ideal} \equiv \Nc \Nf \muq^4/(12 \pi^2)$ with $\muq \equiv \muB / \Nc = \muI / 2$.
We simply evaluate the uncertainty of $P$ by taking the square root of the squared sum of relative errors.
We plot the resulting pressure $P/P_{\rm ideal}$ in Fig.~\ref{fig:ppid}.
The inequality~\eqref{eq:cohen}, rules out the grey hatched region above the lattice data.

We note that the typical value of the normalized pressure $P/P_{\rm ideal}$ inferred from the neutron-star data is less than one.
Also, the pQCD at large $\muB$ predicts $P/P_{\rm ideal} < 1$ as the first coefficient of $\calO(\alpha_s)$ in the perturbative expansion is negative.
By contrast, the normalized pressure in the isospin matter surpasses unity, as can be seen in Fig.~\ref{fig:ppid}.
This clearly indicates that the complex phase in the fermion determinant at the nonzero baryon chemical makes a substantial contribution to reducing the pressure of the baryonic matter.

The exclusion of the high-pressure region in the $\muB - P$ plane can be used to constrain the EoS or the function $P(\varepsilon)$ where $\varepsilon$ is the energy density. It can also be used to constrain the evolution of baryon density $\nB(\muB)$. We will discuss both of these constraints in section \ref{sec:mun}. Here, to gain insight into how the constraints in Fig.~\ref{fig:ppid} translate to constraints on the speed of sound in dense matter, which at zero temperature is defined by the relation $v_s= \sqrt{dP/d\varepsilon}$, we construct simple scenarios in which $v_s$ is constant. In this case, one can use the empirical information about the nuclear saturation point which is characterized by  $P=0$ at  $\muB = \mu_{\rm sat}=923~\text{MeV}$ and $\varepsilon=\varepsilon_{\rm sat} = 150~\text{MeV/fm}^3$ to obtain 
\begin{equation}
    P(\muB) = \frac{v_s^2 \varepsilon_{\rm sat}}{1 + v_s^2} \left[ \left(\frac{\muB}{\mu_{\rm sat}}\right)^{1 + v_s^{-2}} -1\right]\,,
\end{equation}
where $v_s$ is taken to be a constant.
\begin{figure}[ht]
    \centering
    \includegraphics[width=0.95\columnwidth]{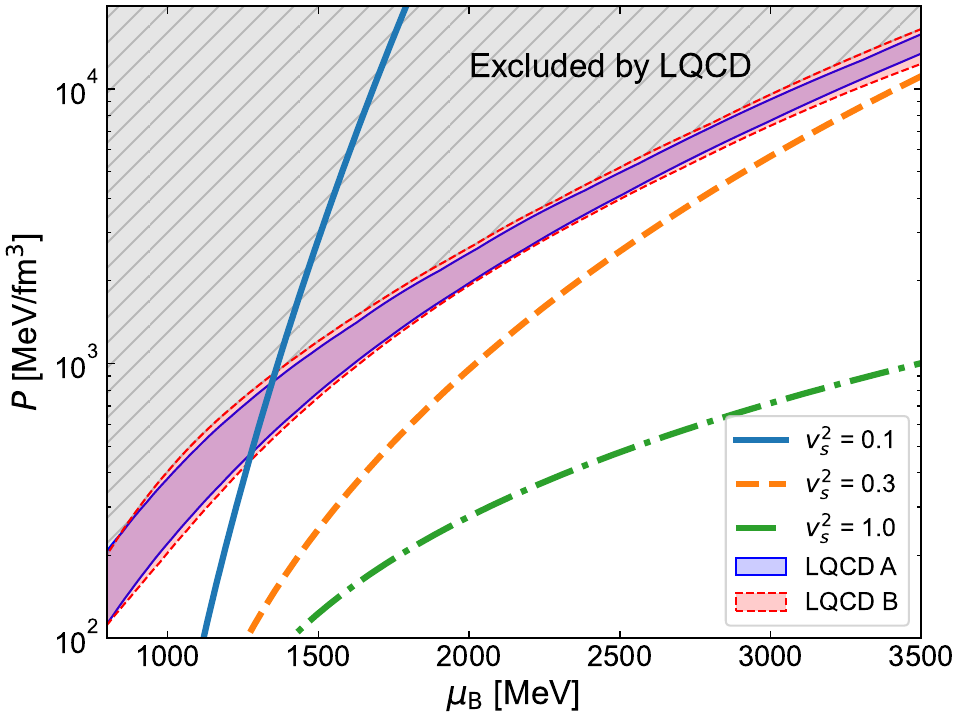}
    \caption{Pressure of the EoS with different values of the constant sound speed.}
    \label{fig:pcss}
\end{figure}
In Fig.~\ref{fig:pcss}, we plot the EoS with different values of $v_s^2$ and see that a softer EoS characterized by a small value of $v_s^2$ has a larger slope in the $P(\muB)$ relation. This can be understood by noting that the sound speed can also be written as
\begin{equation}
\label{eq:vs2}
    v_s^2 = \frac{\nB}{\muB \chi_{\rm B}}\,,
\end{equation}
where $\nB = d P / d\muB$ is the baryon density and $\chi_{\rm B} = d \nB / d\muB$ is the baryon susceptibility;  they correspond to the slope and the curvature of a curve $P(\muB)$, respectively.
As $\muB$ increases, $\chi_{\rm B}$ grows slowly compared to $\nB$ unless an EoS has an extremely soft point such a first-order phase transition, so the stiffness depends dominantly on the value of $\nB$.
The constant extrapolations with small values of the sound speed are excluded by the lattice-QCD constraint as one can see in Fig.~\ref{fig:pcss} that the EoSs with $v_s^2 = 0.1$ are ruled out.
Thus, the upper bound on the function $P(\muB)$ excludes the possibility of having a \emph{soft} EoS over a wide range of $\muB$. This bound on the speed of sound and the average stiffness of the EoS could be employed in modeling heavy-ion collisions where the model assumption about the speed of sound in baryonic is necessary~\cite{Oliinychenko:2022uvy}.

\section{Bounds on $\nB(\muB)$ and $P(\varepsilon)$}
\label{sec:mun}

In this section, we use the integral constraint method developed in Ref.~\cite{Komoltsev:2021jzg} to translate the lattice-QCD constraint on the function $P(\muB)$ to obtain constraints on the functions $\nB(\muB)$ and $P(\varepsilon)$. 
The integral constraint relies on a reference point where all of the thermodynamic properties are known. As mentioned earlier, at low density, the empirical properties of nuclear matter at the saturation density $\nB=\nsat=0.16$ fm$^{-3}$ provides a reference point characterized by $P=0$ at  $\muB = \mu_{\rm sat}=923~\text{MeV}$,  and $\varepsilon=\varepsilon_{\rm sat} = 150~\text{MeV/fm}^3$. At asymptotically high density, one can use the pQCD calculations of the thermodynamic properties to establish a high-density reference point. In what follows, we use the low-density reference point and study its implications. Additional constraints that arise from implementing a high-density reference point will be discussed in section \ref{sec:pQCD}. 

\subsection{Bounds on $\nB(\muB)$}
To establish constraints on the $\muB-\nB$ plane, we first note that thermodynamic consistency requires $P(\muB)$ to be a continuous function and thermodynamic stability requires $(d / d\muB)^2 P(\muB) \geq 0$. 
This implies $d \nB(\muB) / d \muB \geq 0$ and indicates that the function $\nB(\muB)$ cannot decrease with increasing $\muB$.
Further, since causality requires $v_s^2 \leq 1$, Eq.~\eqref{eq:vs2} implies a lower bound on the slope of the function  $\nB(\muB)$ 
\begin{equation}
\label{eq:dnBdmu} 
    \frac{d \nB}{d \muB} \geq \frac{\nB}{\muB}\,.
\end{equation}
Using the low-density reference point, and integrating Eq.~\eqref{eq:dnBdmu} we arrive at a lower bound on the baryon density 
\begin{equation}
\label{eq:nlow}
    n_{\rm min}(\muB) = \frac{n_{\rm sat}}{\mu_{\rm sat}}\muB\,.
\end{equation}

To obtain an upper bound on the baryon density we define a general function $ \check{n}(\muB; \mu_0, n_0)$  to represent all possible behavior of the baryon density $\nB(\muB)$ in the ground state that passes through the point $(\mu_0,n_0)$ and is compatible with Eq.~\eqref{eq:dnBdmu} and subject to the boundary condition set by the low-density reference point. Since $\nB=dP/d\muB$ and $P(\musat)=0$, we obtain the pressure $\check{P}(\muB; \mu_0,n_0)$ associated with $ \check{n}(\muB; \mu_0,n_0)$ at any $\muB>\musat$ by integration, and the QCD inequality in Eq.~\eqref{eq:cohen} reads
\begin{align}
  \label{eq:munbound}
     \check{P}(\muB; \mu_0,n_0) &= \int^{\muB}_{\musat} \!\! d\mu \, \check{n}(\mu; \mu_0, n_0) \,,\nonumber \\
     &\leq \PI(\muI=\frac{2\muB}{\Nc})\,.
\end{align}

\begin{figure}[ht]
    \centering
    \includegraphics[width=0.6\columnwidth]{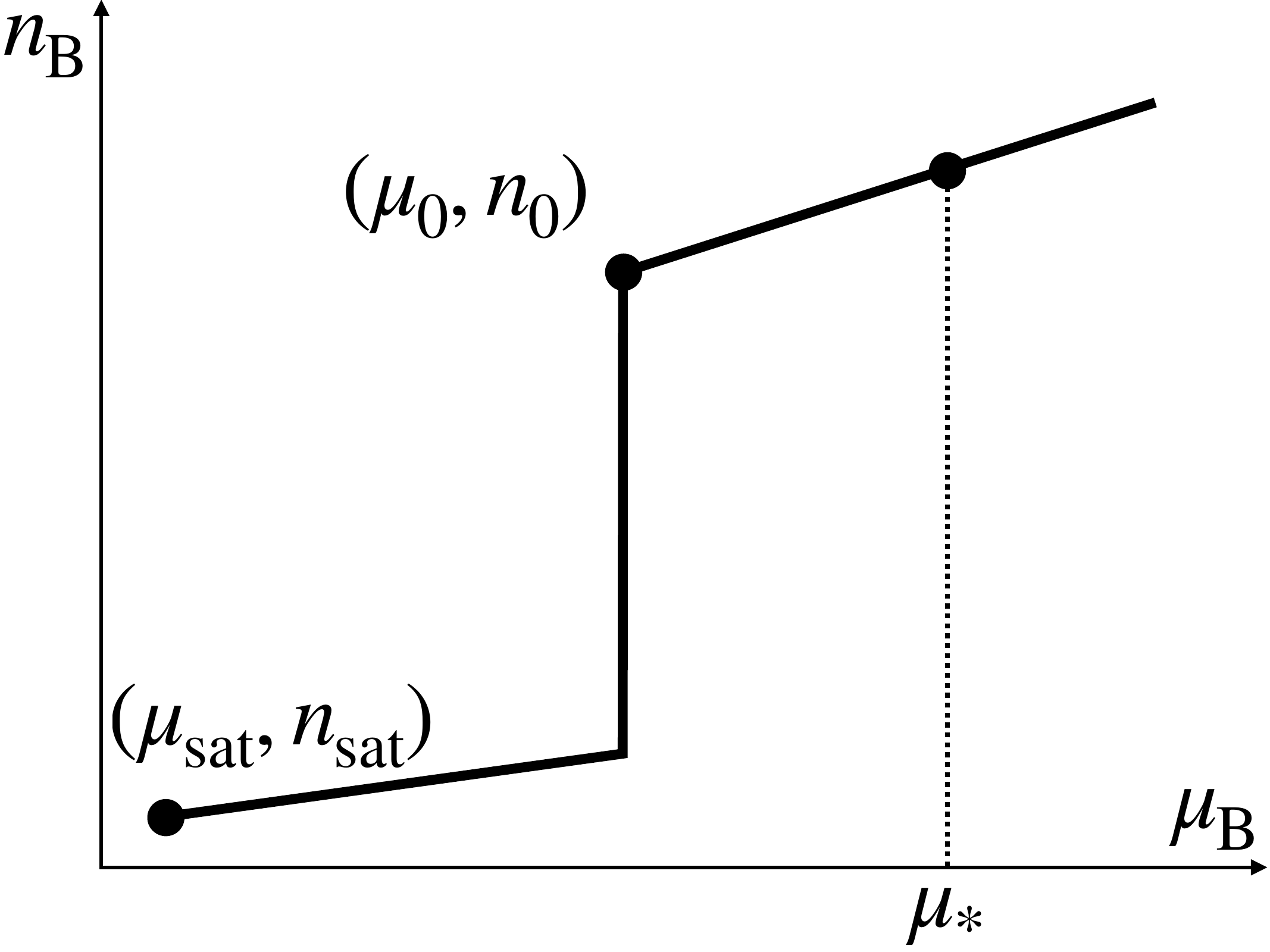}
    \caption{The construction of the baryon density $\check{n}(\muB; \mu_0, n_0)$ such that it minimizes the area at $\mu > \mu_0$.  It extrapolates from the low-density reference point $(\muL, \nL)$, passes through a specific point $(\mu_0, n_0)$, and minimizes the pressure at $\mu \geq \mu_0$.}
    \label{fig:mu0n0_muL}
\end{figure}
To saturate the above inequality, we choose a specific $\check{n}(\muB; \mu_0,n_0)$ that \emph{minimizes} the pressure $\check{P}(\mu; \mu_0,n_0)$ at $\mu$ subject to the low-density reference point. 
This function is shown in Fig.~\ref{fig:mu0n0_muL} and defined as 
\begin{equation}
    \check{n}(\muB; \mu_0,n_0)
    =
    \begin{cases}
        \dfrac{\nsat}{\musat} \muB  & (\musat \leq \muB < \mu_0)\,,\vspace{0.5em}\\
        \dfrac{n_0}{\mu_0} \muB & (\muB \geq \mu_0) \,.
    \end{cases}
\end{equation}

For $\muB < \mu_0$, the baryon density that gives the smallest possible pressure is determined by Eq.~\eqref{eq:nlow} with the smallest slope starting from $(\musat, \nsat)$.
At $\mu_0$, the density jumps to $n_0$ with a first-order phase transition.
Above $\mu_0$, the causal extrapolation from $(\mu_0, n_0)$ sweeps out the smallest area.

We solve the equation $\check{P}(\check{\mu}_\ast; \mu_0, n_0) = \PI(\check{\mu}_\ast)$ for a given $\mu_0$ to find the maximum density $n_0 = n_{\rm  max}(\mu_0)$ comaptible with Eq.~\eqref{eq:cohen}. The point $\check{\mu}_\ast$ is a chemical potential at which $\check{P}$ and $\PI$ intersect.
The solution to this equation gives the maximum density
\begin{equation}
    \label{eq:nmaxlat}
    n_{\rm max}(\muB) = \frac{-\nsat \muB^3  + \muB \musat [\nsat \musat + 2 \PI(\check{\mu}_\ast)]} {\musat (\check{\mu}_\ast^2 - \muB^2)}\,,
\end{equation}
and the location of $\check{\mu}_\ast$ coincides with the tangent point of $\check{P}$ and $\PI$; it is determined by the equation
\begin{equation}
    \label{eq:mucheckast}
    \frac{n_{\rm max}(\muB)}{\muB} \check{\mu}_\ast = \left.\frac{d \PI}{d \muB} \right|_{\muB = \check{\mu}_\ast}\,.
\end{equation}

\begin{figure}[ht]
    \centering
    \includegraphics[width=0.95\columnwidth]{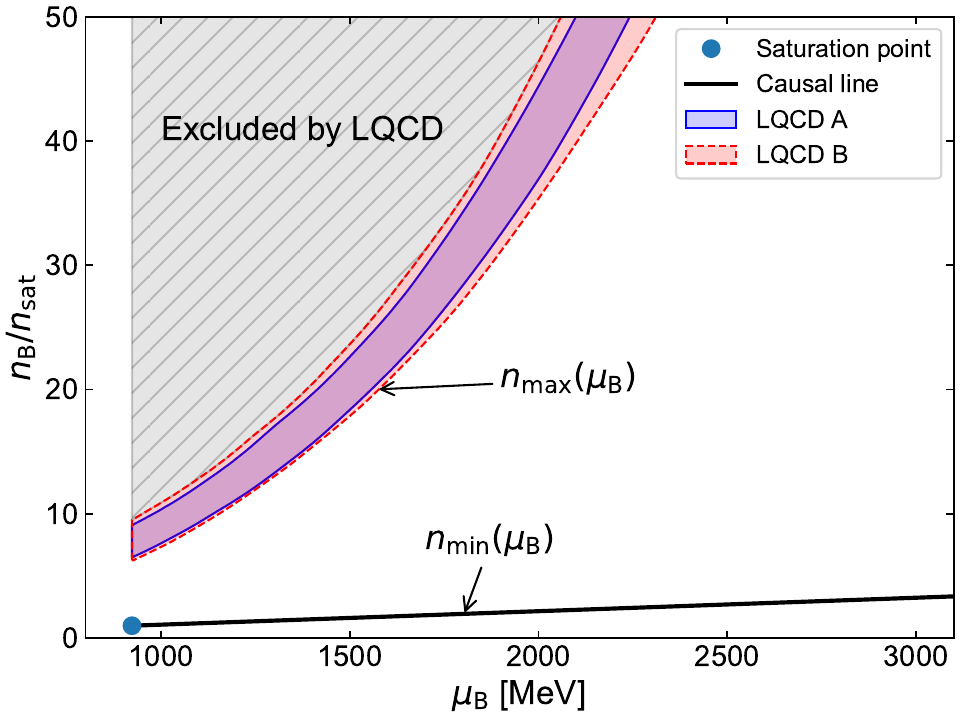}
    \caption{Bound on $\nB(\muB)$ from the lattice-QCD data combined with the saturation property of nuclear matter.}
    \label{fig:mun_NM}
\end{figure}
The lower bound on baryon density defined by Eq.~\eqref{eq:nlow} and the upper bound defined by Eq.~\eqref{eq:nmaxlat} are shown in Fig.~\ref{fig:mun_NM}.
We draw the upper and lower curves of the red and blue bands using the upper and lower bounds on the pressure shown in Fig.~\ref{fig:ppid}, respectively.
We note that the value of $n_{\rm max}(\muB)$ also depend on the slope of $\PI(\muB)$ as is clear from the expression of $\check{\mu}_\ast$~\eqref{eq:mucheckast}, so the red and blue bands shown in Fig.~\ref{fig:mun_NM} may not account for the actual uncertainty of $n_{\rm max}$.
This is also true for the red and blue bands in the figures that appear later.

In deriving $n_{\rm max}(\muB)$, we use the isospin lattice data up to $\muB \simeq 3500~\text{MeV}$.
At $\muB = \muL$, $\check{\mu}_\ast \simeq 1500~\text{MeV}$;  it means that the baryon density around the saturation point is constrained by the isospin lattice data at $\muB \simeq 1500~\text{MeV}$.

The validity range of the lattice bound on the baryon density is limited up to $\check{\mu}_\ast \lesssim 3500~\text{MeV}$ because we use the lattice data only up to $\muB \simeq 3500~\text{MeV}$ so we cannot impose the lattice bound above $\check{\mu}_\ast \gtrsim 3500~\text{MeV}$.
The value of $\check{\mu}_\ast = 3500~\text{MeV}$ is realized at $\muB \simeq 2400~\text{MeV}$ for the ensemble A and $\muB \simeq 2250~\text{MeV}$ for the ensemble B.

We observe that a relation $\check{\mu}_\ast \simeq (3/2) \muB$ holds empirically for a given $\muB$;
it means that the lattice constraint is imposed at $\mu = \check{\mu}_\ast \simeq (3/2) \mu_0$ to put an upper bound on the baryon density at $\mu_0$ in Eq.~\eqref{eq:munbound}.
Meanwhile, the isospin chemical potential $\muI$ of the isospin lattice data is rescaled as $\muB = (3/2) \muI$ to compare them with the baryonic matter.
Therefore they imply that the baryon density at $\muB = \mu_0$ is constrained by the isospin lattice data around $\muI \simeq \mu_0$.

\subsection{Bounds on $P(\varepsilon)$}

\begin{figure}[ht]
    \centering
    \includegraphics[width=0.95\columnwidth]{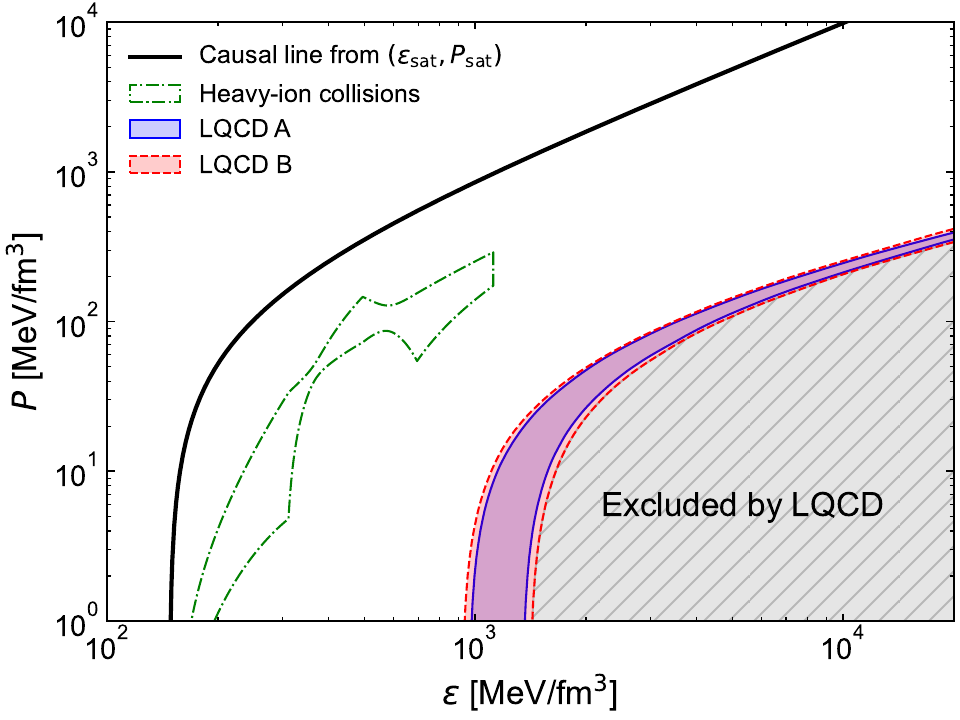}
    \caption{Bound on $P(\varepsilon)$ from the lattice-QCD data combined with the saturation property of nuclear matter.}
    \label{fig:ep_NM}
\end{figure}

Now we translate the bound in the $\muB-\nB$ plane (Fig.~\ref{fig:mun_NM}) to the bound in the $\varepsilon-P$ plane (Fig.~\ref{fig:ep_NM}) following the procedure outlined in Ref.~\cite{Komoltsev:2021jzg}.

To this end, we find the maximum and minimum $\varepsilon$ at a given $\muB$ from the Euler equation $\varepsilon = -P + \muB \nB$ and the isenthalpic condition $h = \varepsilon + P = \muB \nB = \text{const}$.
On the isenthalpic line segment $\varepsilon = -P + h$ in the $\varepsilon-P$ plane, the maximum (minimum) $\varepsilon$ is realized for the minimum (maximum) $P$ on the upper left (lower right) endpoint of the line segment.
Since the maximum and minimum $\varepsilon$ are entangled with the minimum and maximum $P$, we first discuss $P_{\rm min}$ and $P_{\rm max}$ because they can be calculated easily by integrating the $\nB(\muB)$ relation obtained earlier in this section.

At a specific point $(\mu_0, n_0)$ in the $\muB - \nB$ plane, which satisfies the isenthalpic condition $n_0 = h / \mu_0$, the minimum pressure is given by the integration of $n_{\rm min}$ followed by the first-order phase transition at $\mu_0$
\begin{equation}
  P_{\rm min}(\mu_0) = \frac{\nL}{2 \muL} \left(\mu_0^2 -  \muL^2 \right)\,.
\end{equation}
Note that the minimum pressure depends only on $\mu_0$ but not on $n_0$ and $h$, so the pressure takes the smallest value at the smallest possible $\mu_0$.
Such $\mu_0$ is realized at the intersection of the isenthalpic line $\nB = h/\muB$ and the maximum density $\nB = n_{\rm max}(\muB)$ in the $\muB-\nB$ plane.

Likewise, the maximum pressure at a specific point $(\mu_0, n_0)$, which satisfies the isenthalpic condition $n_0 = h / \mu_0$, is
\begin{align}
  &P_{\rm max}(\mu_0, n_0 = h/\mu_0) \notag \\
  &=
    \begin{cases}
      \frac{h}{2}\left(1 - \frac{\muL^2}{\mu_0^2}\right) & (n_0 \leq \frac{n_{\rm max}(\muL)}{\muL} \mu_0)\,,\\
    \frac{h}{2} \left(1 - \frac{\mu_u^2}{\mu_0^2}\right)
    + \int_{\muL}^{\mu_{\rm u}} \!\! d\mu'\, n_{\rm max}(\mu')  & (n_0 > \frac{n_{\rm max}(\muL)}{\muL} \mu_0)\,,
  \end{cases}
\end{align}
where the upper bound of the integral in the latter case, $\mu_{\rm u}$, is the intersection of the line $\nB = (n_0 / \mu_0) \muB$ with the curve $\nB = n_{\rm max}(\muB)$,
\begin{equation}
    \mu_{\rm u} = \sqrt{\frac{\muL [\check{\mu}_\ast^2 n_0 - \mu_0 \muL \nL - 2\mu_0 \PI(\check{\mu}_\ast)]}
    {\muL n_0 - \mu_0 \nL}}\,.
\end{equation}
From the above expression, the maximum pressure $P_{\rm max}$ takes the largest value at the largest possible $\mu_0$, which is realized at the intersection of the isenthalpic line $\nB = h/\muB$ and the minimum density $\nB = n_{\rm min}(\muB)$ in the $\muB-\nB$ plane.

The upper end of the isenthalpic line segment in the $\varepsilon-P$ plane is
\begin{equation}
\label{eq:isenthalpy_up}
    \begin{pmatrix}
        \varepsilon \\ P
    \end{pmatrix}
    =
    \begin{pmatrix}
    h-P_{\rm max}\{\mu_{\rm max}(h), n_{\rm min}[\mu_{\rm max}(h)]\} \\
    P_{\rm max}\{\mu_{\rm max}(h), n_{\rm min} [\mu_{\rm max}(h)]\}
    \end{pmatrix}\,,
\end{equation}
where $\mu_{\rm max}(h)$ is given by the intersection of $\nB = h / \muB$ and the $\nB = n_{\rm min}(\muB)$.
The lower end is
\begin{equation}
\label{eq:isenthalpy_low}
    \begin{pmatrix}
        \varepsilon \\
        P
    \end{pmatrix}
    =
    \begin{pmatrix}
        h-P_{\rm min}[\mu_{\rm min}(h)]\\
        P_{\rm min}[\mu_{\rm min}(h)]
    \end{pmatrix}\,,
\end{equation}
where $\mu_{\rm min}(h)$ is given by the intersection of $\nB = h / \muB$ and the $\nB = n_{\rm max}(\muB)$.
By substituting $h = \muB n_{\rm min}(\muB)$ in Eq.~\eqref{eq:isenthalpy_up} and $h = \muB n_{\rm max}(\muB)$ in Eq.~\eqref{eq:isenthalpy_low}, we find the upper and the lower bound on the allowed range of values in the $\varepsilon-P$ plane as parametric equations with $\muB$ as a parameter.
The parametric equation for the upper bound is
\begin{equation}
\label{eq:ep_up}
    \begin{pmatrix}
        \varepsilon \\
        P
    \end{pmatrix}
    = 
    \begin{pmatrix}
        \varepsilon_{\rm min}(\muB)\\
        P_{\rm max}[\muB, n_{\rm min} (\muB)]
    \end{pmatrix}\,,
\end{equation}
and that for the lower bound is
\begin{equation}
\label{eq:ep_low}
    \begin{pmatrix}
        \varepsilon \\
        P
    \end{pmatrix}
    =
    \begin{pmatrix}
    \varepsilon_{\rm max}(\muB)\\
    P_{\rm min}(\muB)
    \end{pmatrix}\,,
\end{equation}
where the minimum and maximum energy densities are defined as
\begin{equation}
\label{eq:eminmax}
\begin{split}
    \varepsilon_{\rm min}(\muB) &= - P_{\rm max}[\muB, n_{\rm min}(\muB)] + \muB n_{\rm min}(\muB)\,,\\
    \varepsilon_{\rm max}(\muB) &= - P_{\rm min}(\muB) + \muB n_{\rm max}(\muB)\,.
\end{split}
\end{equation}

In Fig.~\ref{fig:ep_NM}, we plot the bound in the $\varepsilon-P$ plane.
The lattice-QCD data constrains the soft part of the EoS as explained in the previous section.
The upper bound matches with the causal extrapolation from the point $(\varepsilon_{\rm sat}, \PL)$ with $v_s^2=1$.
The heavy-ion constraint from the hadron transport model is also overlaid~\cite{Oliinychenko:2022uvy}.

\section{Combining lattice-QCD data with pQCD reference point}
\label{sec:pQCD}

In this section, we use the pQCD information in addition to the lattice data of isospin QCD matter and the empirical saturation property of nuclear matter;  we discuss the modification to the bound on $\nB(\muB)$ and $P(\varepsilon)$.

\subsection{High-density reference point from perturbative QCD}
\label{sec:muH}

For the pQCD thermodynamics, we use the result expanded up to $\calO(\alpha_s^2)$~\cite{Freedman:1976xs, *Freedman:1976dm, *Freedman:1976ub, Baluni:1977ms} in the $\overline{\rm MS}$ scheme~\cite{Fraga:2001id, Kurkela:2009gj} for the massless $\Nf=2$ quarks.
We use the perturbative coefficients concisely summarized in Table.~II of Ref.~\cite{Gorda:2023mkk}.
We assume the running of $\alpha_s(\bar{\Lambda})$ at the N2LO and take its scale as $\bar{\Lambda} = 2 \muB / \Nc$.
The $\overline{\rm MS}$ scale is fixed as $\Lambda_{\overline{\rm MS}} \simeq 330~\text{MeV}$, which is the value suggested from the $\Nf=2$ lattice-QCD data~\cite{Fritzsch:2012wq, FlavourLatticeAveragingGroupFLAG:2021npn}.
The uncertainty corresponding to the ambiguity in the choice of $\bar{\Lambda}$ is commonly evaluated in the literature by varying it by a factor of two, namely taking $X\equiv \bar{\Lambda} / (2 \muB / \Nc)$ as $X \in [1/2, 2]$;
here we also follow this convention.

\begin{table}[ht]
    \centering
    \begin{tabular}{c | c | c}
        $\muH$ [MeV] & $\nH$ [$\nsat$] & $\PH$ [MeV/fm$^3$] \\ \hline
        3000 & $43.86^{+ 1.86}_{- 2.47}$ & $4982^{+ 353}_{- 882}$
    \end{tabular}
    \caption{The high-density reference points from the pQCD thermodynamics.  The uncertainties arises from the ambiguity in the choice of the renormalization scale $\bar{\Lambda}$, which is taken to be $\bar{\Lambda} = 2 \muH / \Nc$ and varied by a factor 2.}
    \label{tab:pQCD}
\end{table}
We choose the high-density reference point $(\muH, \nH, \PH)$ as tabulated in Table.~\ref{tab:pQCD}.
Throughout this work, we fix $\muH $ as $3000~\text{MeV}$ although one may be able to push down $\muH$ to $2700~\text{MeV}$ as this value achieves the relative scale variation uncertainty of $\sim 24~\%$, which is the standard value used in the literature as in Refs.~\cite{Kurkela:2014vha, Annala:2017llu, Annala:2019puf, Annala:2021gom, Komoltsev:2021jzg}.

\subsection{Bounds on $\nB(\muB)$}

Combining the causal extrapolation from the high-density reference point $\nB = (\nH / \muH) \muB$ and the lattice upper bound~\eqref{eq:nmaxlat} obtained in Sec.~\ref{sec:mun}, the maximum density is modified as
\begin{align}
    & n_{\rm max}(\muB) \notag \\
    & =
    \begin{cases}
        \frac{- \nL \muB^3  + \muB \muL [\nL \muL + 2 \PI(\check{\mu}_\ast)]}{\muL (\check{\mu}_\ast^2 - \muB^2)} & (\muL \leq \muB < \hat{\mu}_{\rm c})\,, \\
        \frac{\nH}{\muH} \muB & (\hat{\mu}_{\rm c} \leq \muB \leq \muH)\,,
    \end{cases}
    \label{eq:nmax}
\end{align}
where $\hat{\mu}_{\rm c}$ is given by the intersection of the above two cases
\begin{equation}
    \hat{\mu}_{\rm c} = \sqrt{\frac{\muL [\check{\mu}_\ast^2 \nH - \muH \muL \nL - 2\muH \PI(\check{\mu}_\ast) ]}
    {\muL \nH - \muH \nL}}\,.
\end{equation}

With the high-density reference point, the lower bound on the baryon density is also subject to the lattice bound.
To discuss a modification to the lower bound~\eqref{eq:nlow}, we define a general function $\hat{n}(\muB; \mu_0, n_0)$ to represent all possible configuration of the baryon density $\nB(\muB)$ in the ground state that passes through the point $(\mu_0, n_0)$ and is subject to the causality and the boundary condition set by the high-density reference point.
We obtain the pressure $\hat{P}(\muB; \mu_0, n_0)$ corresponding to $\hat{n}(\muB; \mu_0, n_0)$ at any $\muB < \muH$ by integration;
the QCD inequality~\eqref{eq:cohen} reads
\begin{align}
  \hat{P}(\muB; \mu_0, n_0) &= \PH - \int_{\muB}^{\muH} \!\! d\mu \, \hat{n}(\mu; \mu_0, n_0)\notag \\
  & \leq \PI (\muI = \frac{2\muB}{\Nc})\,.
  \label{eq:munbound_pqcd}
\end{align}

\begin{figure}[ht]
    \centering
    \includegraphics[width=0.6\columnwidth]{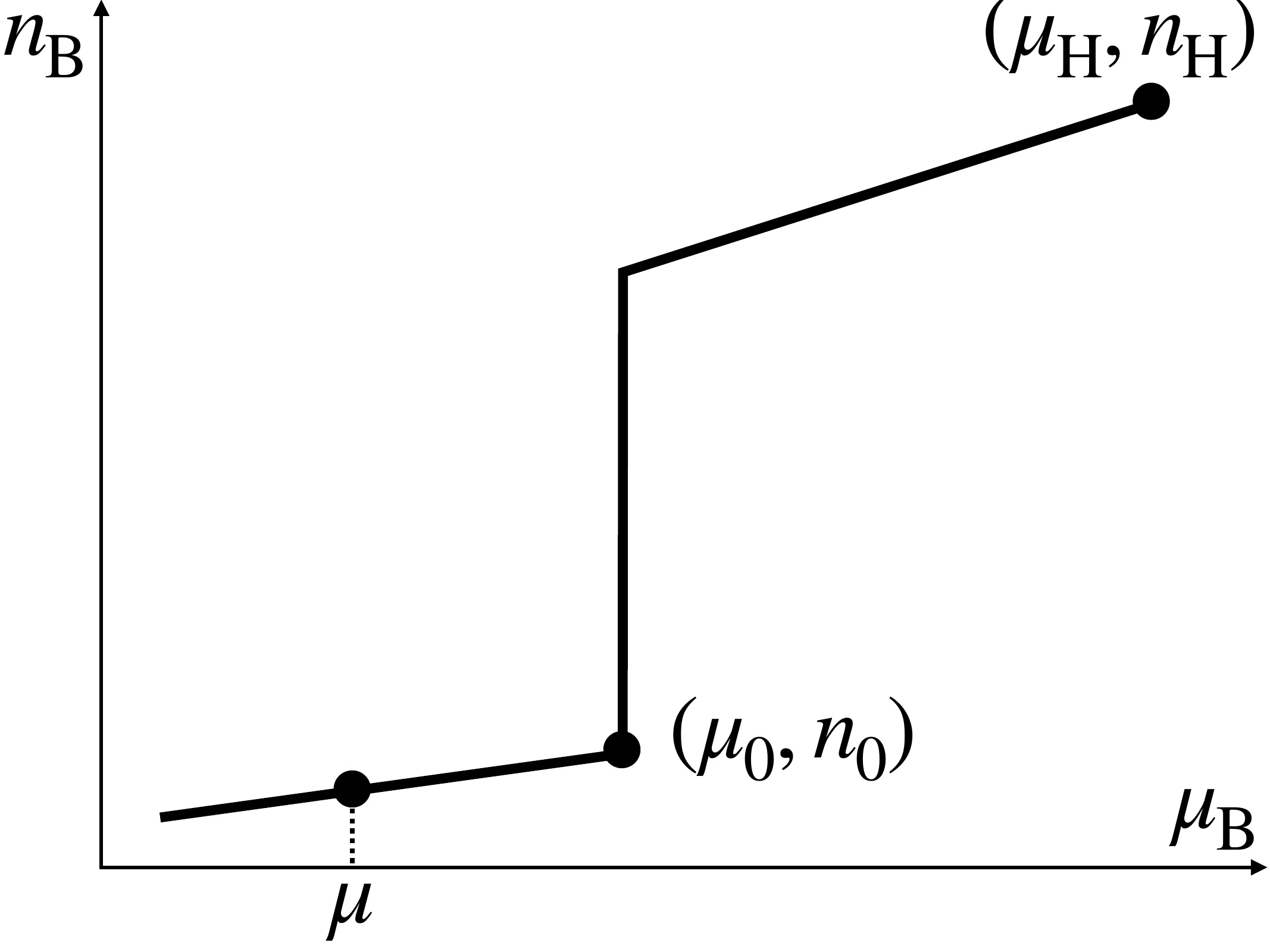}
    \caption{The construction of the baryon density $\hat{n}(\muB; \mu_0, n_0)$ such that it maximizes the area at $\mu < \mu_0$.  It extrapolates from the high-density reference point $(\muH, \nH)$, passes through a specific point $(\mu_0, n_0)$, and minimizes the pressure $\hat{P}(\mu; \mu_0, n_0)$ at $\mu \geq \mu_0$.}
    \label{fig:mu0n0_muH}
\end{figure}
To saturate the above inequality, we choose a specific $\hat{n}(\muB; \mu_0, n_0)$ that \emph{minimizes} the pressure $\hat{P}(\mu; \mu_0, n_0)$ at $\mu$ subject to the high-density reference point.
This is equivalent to \emph{maximizing} the area beneath $\hat{n}(\muB; \mu_0, n_0)$.
This function is shown in Fig.~\ref{fig:mu0n0_muH} and defined as
\begin{equation}
    \hat{n}(\muB; \mu_0, n_0)
    =
    \begin{cases}
        \dfrac{n_0}{\mu_0} \muB  & (\muB < \mu_0) \,,\vspace{0.5em}\\
        \dfrac{\nH}{\muH} \muB & (\mu_0 \leq \muB \leq \muH)\,.
    \end{cases}
\end{equation}

For $\muB < \mu_0$, the baryon density that sweeps out the largest area is the causal extrapolation with the largest slope starting from $(\mu_0, n_0)$.
At $\mu_0$, the density jumps from $n_0$ to $(\nH / \muH) \mu_0$ with a first-order phase transition.
Above $\mu_0$, the baryon density that sweeps out the largest possible area is determined by the latter case of Eq.~\eqref{eq:nmax} with the causal extrapolation from $(\muH, \nH)$.

We solve the equation $\hat{P}(\hat{\mu}_\ast; \mu_0, n_0) = \PI(\hat{\mu}_\ast)$ for a given $\mu_0$ to find a minimum density $n_0 = n_{\rm min}(\mu_0)$ compatible with Eq.~\eqref{eq:cohen}.
The point $\hat{\mu}_\ast$ is a chemical potential at which $\hat{P}$ and $\PI$ intersect.
The solution to this equation gives the lattice-QCD lower bound on the density
\begin{equation}
    \label{eq:nminlat}
    n_0(\mu_0) = \frac{\nH \mu_0^3  - \mu_0 \muH \{\nH \muH - 2 [\PH - \PI(\hat{\mu}_\ast)]\}}{\muH (\mu_0^2 - \hat{\mu}_\ast^2)}\,,
\end{equation}
and the location of $\hat{\mu}_\ast$ coincides with the tangent point of $\hat{P}$ and $\PI$; it is defined by the equation
\begin{equation}
    \label{eq:muhatast}
    \frac{n_0(\mu_0)}{\mu_0} \hat{\mu}_\ast = \left.\frac{d \PI}{d \muB} \right|_{\muB = \hat{\mu}_\ast}\,.
\end{equation}
So far, we have not used the information of the low-density reference point $(\muL, \nL)$.
We combine the lattice bound~\eqref{eq:nminlat} with the causal extrapolation from $(\muL, \nL)$~\eqref{eq:nlow}, we obtain the minimum density
\begin{align}
    & n_{\rm min}(\muB) \notag \\
    & =
    \begin{cases}
        \frac{\nL}{\muL} \muB & (\muL \leq \muB < \check{\mu}_{\rm c})\,, \\
        \frac{\nH \muB^3  - \muB \muH \{\nH \muH - 2 [\PH - \PI(\hat{\mu}_\ast)]\}}{\muH (\muB^2 - \hat{\mu}_\ast^2)} & (\check{\mu}_{\rm c} \leq \muB \leq \muH)\,,
    \end{cases}
    \label{eq:nmin}
\end{align}
where $\check{\mu}_{\rm c}$ is given by the intersection of the above two cases, namely, the causal line and the lattice bound,
\begin{equation}
    \check{\mu}_{\rm c} = \sqrt{\frac{\muH \{\muL \muH \nH - \hat{\mu}_\ast^2 \nL - 2\muL [\PH - \PI(\hat{\mu}_\ast)]  \}}
    {\muL \nH - \muH \nL}}\,.
\end{equation}

\begin{figure}[ht]
    \centering
    \includegraphics[width=0.95\columnwidth]{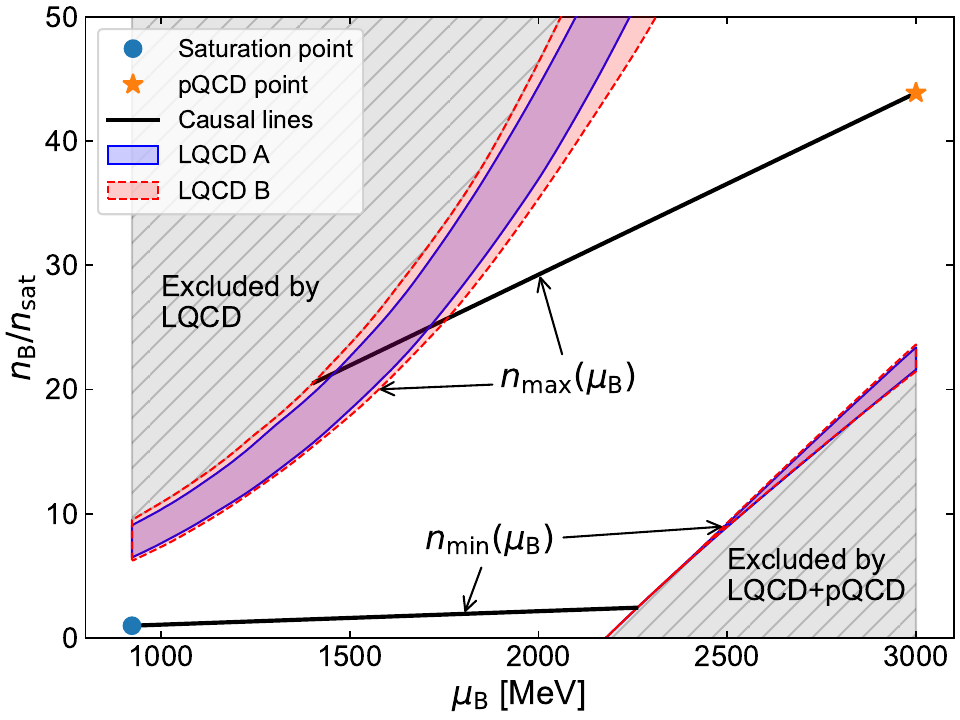}
    \caption{Bound on $\muB(\nB)$ from the lattice-QCD data combined with the high-density reference point calculated from pQCD.}
    \label{fig:mun_pQCD}
\end{figure}
The upper bound on baryon density defined by Eq.~\eqref{eq:nmax} and the lower bound defined by Eq.~\eqref{eq:nmin} are shown in Fig.~\ref{fig:mun_pQCD}.
The red and blue bands are the bounds obtained from the lattice data and the black lines correspond to the causal extrapolations from the low- and high-density reference points.

The validity range of the lattice bound on the baryon density is limited above $\hat{\mu}_\ast \gtrsim 270~\text{MeV}$ because we use the lattice data only above $\muB \simeq 270~\text{MeV}$ so we cannot impose the lattice bound below $\hat{\mu}_\ast \lesssim 270~\text{MeV}$.
The value of $\check{\mu}_\ast = 270~\text{MeV}$ is realized at $\muB \simeq 2200~\text{MeV}$ for $\muH = 3000~\text{MeV}$.

We find an empirical relation $\hat{\mu}_\ast + \muH \simeq (3/2) \muB$ for a given $\muB$;
it means that the lattice constraint is imposed at $\mu = \hat{\mu}_\ast \simeq (3/2) \mu_0 - \muH$ to put an upper bound on the baryon density at $\mu_0$ in Eq.~\eqref{eq:munbound_pqcd}.
This imply that when combined with the pQCD data imposed at $\muH$, the baryon density at $\muB = \mu_0$ is constrained by the isospin lattice data around $\muI \simeq \mu_0 - (2/3) \muH$.

\subsection{Bounds on $P(\varepsilon)$}
\begin{figure}[ht]
    \centering
    \includegraphics[width=0.95\columnwidth]{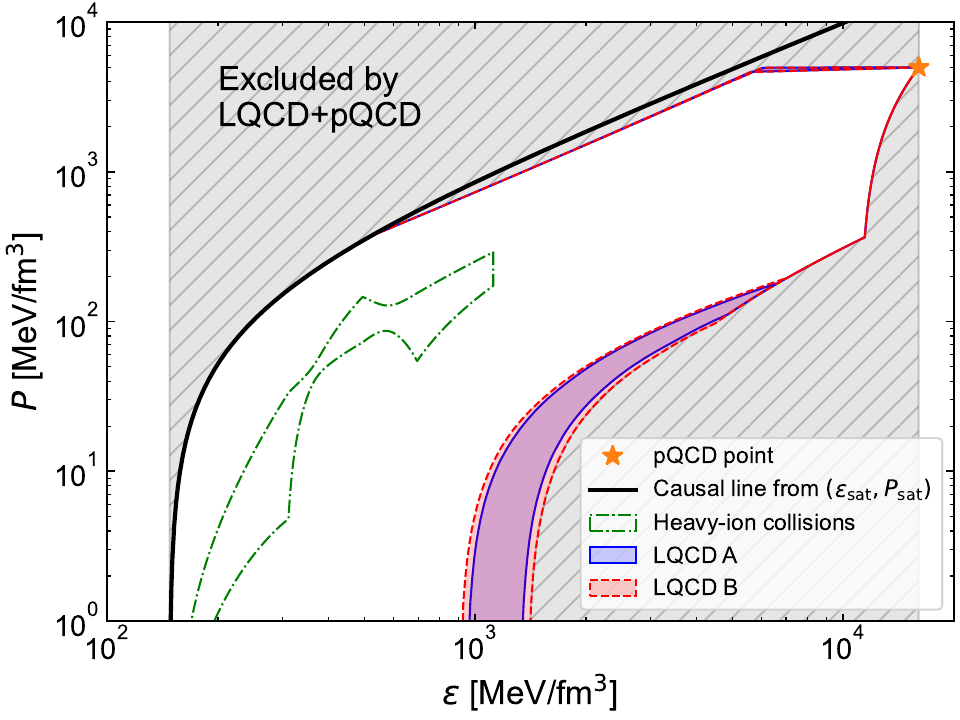}
    \caption{Bound on $P(\varepsilon)$ from the lattice-QCD data combined with the high-density reference point calculated from pQCD.}
    \label{fig:ep_pQCD}
\end{figure}

Now we translate the bound in the $\muB-\nB$ plane (Fig.~\ref{fig:mun_pQCD}) to the bound in the $\varepsilon-P$ plane (Fig.~\ref{fig:ep_pQCD}) following the procedure outlined in the previous section.
The parametric equation for the upper bound is Eq.~\eqref{eq:ep_up} and the equation for the lower bound is Eq.~\eqref{eq:ep_low}.

The only modification occurs in the expression of $P_{\rm min}$.
The minimum pressure at $(\mu_0, n_0)$ is
\begin{align}
    &P_{\rm min}(\mu_0) \notag\\
    &=
    \begin{cases}
    \frac{\nL}{2 \muL} (\mu_0^2 - \muL^2) & (\muL \leq \mu_0 < \tilde{\mu}_{\rm c})\,, \\
    \PH - \frac{\nH}{2\muH}(\muH^2 - \mu_0^2) & (\tilde{\mu}_{\rm c} \leq \mu_0 \leq \muH) \,,
    \end{cases}
    \label{eq:Pmin}
\end{align}
where $\tilde{\mu}_{\rm c}$ is defined as
\begin{equation}
    \tilde{\mu}_{\rm c} = \sqrt{\frac{\muL\muH(\muH \nH - \muL \nL -2 \PH )}{\muL \nH - \muH \nL}}\,.
\end{equation}
Remember that the minimum pressure does not depend on $n_0$.
The former case in Eq.~\eqref{eq:Pmin} is given by $\int_{\muL}^{\mu_0} d\mu\, n_{\rm min}(\mu)$ while the latter case is given by $\PH - \int_{\mu_0}^{\muH} d\mu\, n_{\rm max}(\mu)$.
Since $\hat{\mu}_{\rm c} < \tilde{\mu}_{\rm c} < \check{\mu}_{\rm c}$, the integrals of $n_{\rm min}$ and $n_{\rm max}$ are carried out straightforwardly.
We then arrive at the expression as simple as Eq.~\eqref{eq:Pmin}.

In Fig.~\ref{fig:ep_pQCD}, we plot the upper and lower bounds in the $\varepsilon-P$ plane that are subject to the high-density reference point.
The upper and lower bounds are defined in Eq.~\eqref{eq:ep_up} and \eqref{eq:ep_low}, respectively.
We observe that the lattice-QCD data now constrains the stiff part of the EoS in addition to the soft part of the EoS by including the high-density reference point in the integral.
Further, the lower bound is also modified as we require the EoS to converge to the high-density point on the $\varepsilon - P$ plane.

\section{Comparison to the pQCD integral constraint}
\label{sec:integ}

In this section, we compare the lattice-QCD constraint with the constraint put by the thermodynamically consistent construction of the EoS imposing the integral condition:
\begin{equation}
    \label{eq:pQCDinteg}
    \int_{\muL}^{\muH} d\mu' \nB(\mu') = \PH - \PL\,.
\end{equation}
In the following, we loosely refer this constraint to as the ``pQCD integral constraint''.

The minimum density from the pQCD integral constraints is~\cite{Komoltsev:2021jzg}
\begin{align}
    & n_{\rm min}^{\rm pQCD}(\muB) \notag \\
    &=
    \begin{cases}
        \frac{\nL}{\muL} \muB & (\muL \leq \muB < \mu_{\rm c}^{\rm pQCD})\,,\\
         \frac{\nH \muB^3 - \muB \muH (\nH \muH - 2 \PH)}{\muH (\muB^2 - \muL^2)} & (\mu_{\rm c}^{\rm pQCD} \leq \muB \leq \muH)\,,
    \end{cases}
    \label{eq:nminpqcd}
\end{align}
and the maximum density is
\begin{align}
    & n_{\rm max}^{\rm pQCD}(\muB) \notag \\
    &=
    \begin{cases}
         \frac{- \nL \muB^3 + \muB \muL (\nL \muL + 2 \PH )}{\muL (\muH^2 - \muB^2)} & (\muL \leq \muB < \mu_{\rm c}^{\rm pQCD})\,,\\
        \frac{\nH}{\muH} \muB & (\mu_{\rm c}^{\rm pQCD} \leq \muB \leq \muH)\,,
    \end{cases}
    \label{eq:nmaxpqcd}
\end{align}
where $\mu_{\rm c}^{\rm pQCD}$ is
\begin{equation}
    \mu_{\rm c}^{\rm pQCD} = \sqrt{\frac{\muL\muH(\muH \nH - \muL \nL -2 \PH )}{\muL \nH - \muH \nL}}\,.
\end{equation}
The maximum pressure at a given $\muB$ constructed consistently with the constraint~\eqref{eq:pQCDinteg} is
\begin{equation}
    \label{eq:Pmaxpqcd}
    P_{\rm max}^{\rm pQCD}(\muB) = \PH \frac{\muB^2 - \muL^2}{\muH^2 - \muL^2}\,.
\end{equation}

\begin{figure}[ht]
    \centering
    \includegraphics[width=0.95\columnwidth]{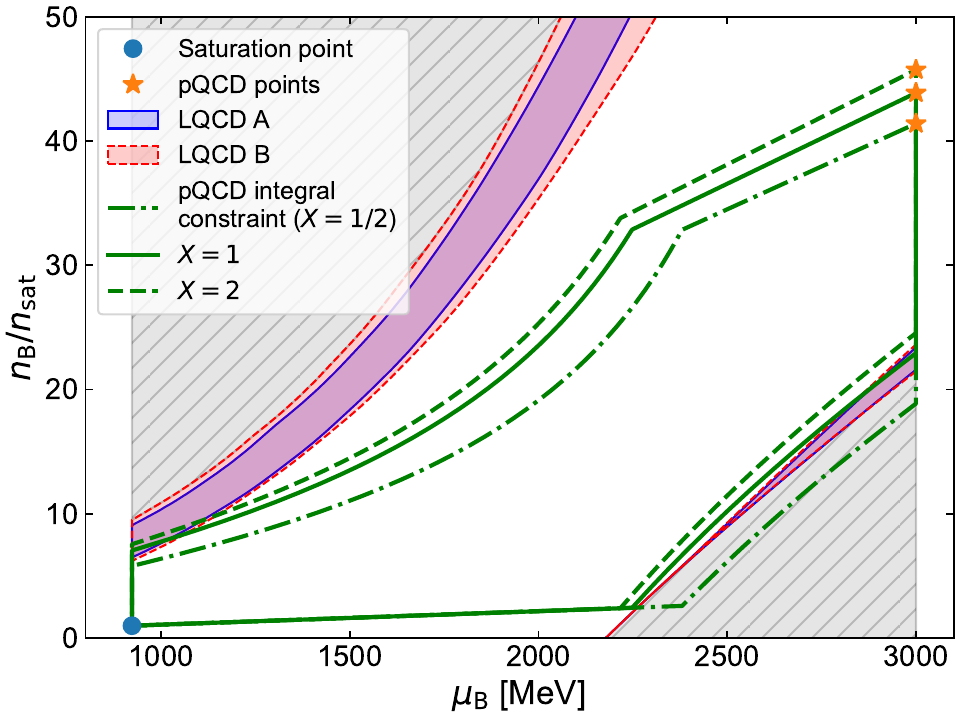}
    \caption{Comparison of the lattice bound on $\nB(\muB)$ relation and the pQCD integral constraint.  For the pQCD integral constraint, we also take into account the renormalization scale ambiguity by varying by a factor of two.}
    \label{fig:mun_integ}
\end{figure}
\begin{figure}[ht]
    \centering
    \includegraphics[width=0.95\columnwidth]{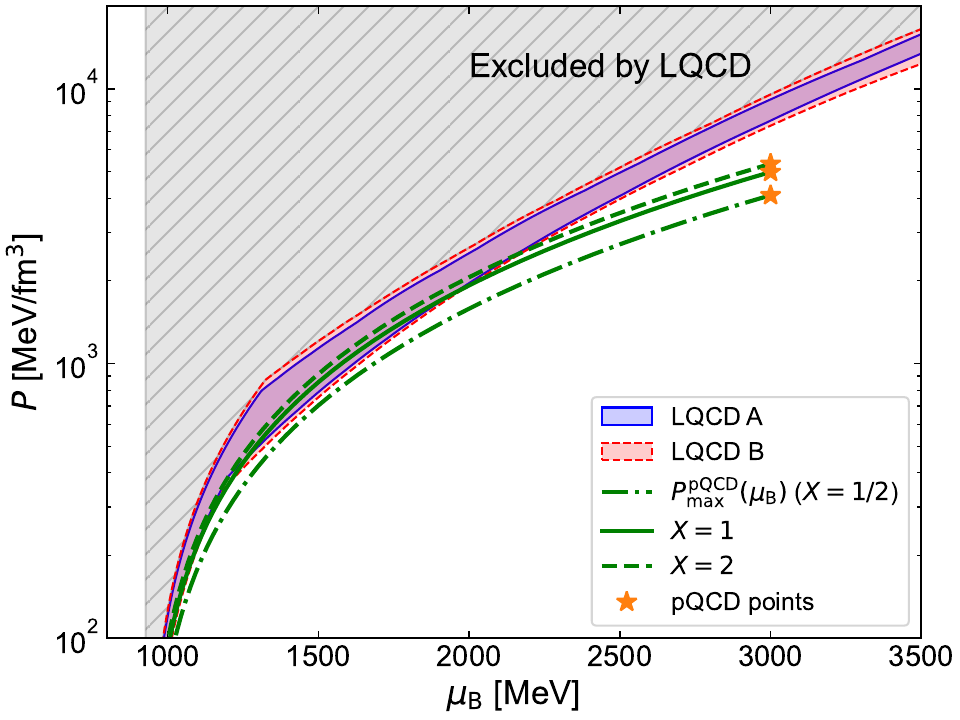}
    \caption{Comparison of the lattice bound on $P(\muB)$ relation taking into account the saturation property and the maximum pressure from the pQCD integral constraint.}
    \label{fig:pmu_integ}
\end{figure}
In Fig.~\ref{fig:mun_integ}, we plot the pQCD integral constraint in the $\muB-\nB$ plane along with the lattice constraint.
From the figure, we can tell that the lattice bound can constrain better around $\muL$ and $\muH$.
In the pQCD integral constraint, the effect of the scale ambiguity is also included by choosing the factor $X = 1/2$, $1$, and $2$ as explained in Sec.~\ref{sec:muH}.
We note that the lower bound from the lattice data in Fig.~\ref{fig:mun_integ} also varies according to the choice of $X$, but we do not include these effects here to make the figure simple.
The upper bound in Fig.~\ref{fig:mun_integ} does not vary;  the only source of the uncertainty for this bound is the lattice errors.

In Fig.~\ref{fig:pmu_integ}, we plot the maximum pressure~\eqref{eq:Pmaxpqcd} from the pQCD integral constraint, and we take the effect of the scale ambiguity into account as in Fig.~\ref{fig:mun_integ}.
These green lines in the figure are compatible with both the low-density saturation and high-density pQCD reference points, and thus can be considered as the pressure upper bound in such a setup.
For the lattice constraint, we incorporate the empirical information on the nuclear saturation, i.e., the pressure vanishes at $\musat$.
We integrate $n_{\rm max}(\muB)$~\eqref{eq:nmaxlat} to include such an effect and combine it with the bare lattice data presented in Fig.~\ref{fig:pcss}.
Around $\muB \simeq 1500~\text{MeV}$, the lattice bound is as constraining as the pQCD bound.

The pQCD integral constraint becomes more constraining compared to the lattice bound when we take smaller value for $\muH$ and vice versa when we take large $\muH$.
Both constraints have different sources of uncertainty, so the comparison will lead to an independent check of each constraint.
Furthermore, in addition to that the independent check is feasible, we can also benefit from having two independent constraints as we can put improved bounds by combining these two.

In what follows, we outline how improved bounds can be obtained from the synergy of the pQCD and lattice-QCD constraints.
We can simply obtain the improved bounds by taking the more restrictive one out of the lattice bound and the pQCD integral constraint.
For instance, if we compare the lower curve of the band of the lattice upper bound with the pQCD integral constraint with $X=2$ in Fig.~\ref{fig:mun_integ}, the former is more restrictive around $\muB = 1000~\text{MeV}$.
So, the improved bound in this case is patching the lattice bound around $\muB \lesssim 1000~\text{MeV}$ and the pQCD integral constraint at $\muB \simeq 1000~\text{MeV}$.
The same construction works for Fig.~\ref{fig:pmu_integ}.

In Fig.~\ref{fig:mun_integ} and Fig.~\ref{fig:pmu_integ}, the range of $\muB$ at which the lattice bound is more restrictive compared to the pQCD bound is different.
To understand this difference, we compare the semi-analytic formulae for the lattice constraint and the pQCD integral constraint.
As a particular example, we compare the maximum density in the lattice constraint~\eqref{eq:nmax} and that in the pQCD integral constraint~\eqref{eq:nmaxpqcd} around $\musat$.
The lattice maximum density is obtained by replacing $(\muH, \PH)$ with $(\check{\mu}_\ast, \PI(\check{\mu}_\ast))$ in the former case of the pQCD maximum density~\eqref{eq:nmaxpqcd}.
At $\muB=\muL$, the maximum density from the lattice and the pQCD constraint are
\begin{equation}
\begin{split}    
    n_{\rm max}^{\rm lat}(\muL) &= \frac{2\muL \PI(\check{\mu}_\ast) }{\check{\mu}_\ast^2 - \muL^2}\,,\\
    n_{\rm max}^{\rm pQCD}(\muL) &= \frac{2\muL \PH }{\muH^2 - \muL^2}\,.
\end{split}
\end{equation}
At $\muL$, $\check{\mu}_\ast \simeq 1500~\text{MeV}$.
In Fig.~\ref{fig:pmu_integ}, we observe that
\begin{equation}
    P_{\rm max}^{\rm pQCD}(\check{\mu}_\ast) \gtrsim \PI(\check{\mu}_\ast)\,.
\end{equation}
Remember that here we compare the lower curve of the lattice band and the $X=2$ of the pQCD curves.
By using the relation~\eqref{eq:Pmaxpqcd}, it leads that the lattice bound is more restrictive at $\musat$, i.e. $n_{\rm max}^{\rm pQCD} \gtrsim n_{\rm max}^{\rm lat}(\musat)$.
So, even though the range of $\muB$ at which the lattice bound is more restrictive compared to the pQCD bound is different in the $\muB - \nB$ and the $\muB - P$ plane, they are consistent with each other from the discussion above.
Thus, we can safely patch together the lattice and pQCD bounds on the $\nB(\muB)$ and $P(\muB)$ relations at different values of $\muB$.

\section{Summary and conclusion}

We demonstrated that the equation of state of two-flavor symmetric matter at nonzero baryon chemical potential (i.e.\ the symmetric nuclear matter) can be robustly constrained by combining a QCD inequalities~\cite{Cohen:2003ut, *Cohen:2004qp} and the recent calculation of the equation of state of matter at nonzero isospin chemical potential on the lattice~\cite{Abbott:2023coj}. We presented the lattice constraints in three ways:
(a) the bound on the pressure at a given baryon chemical potential (Figs.~\ref{fig:ppid}, \ref{fig:pcss}),
(b) the bounds on the baryon density at a given baryon chemical potential (Figs.~\ref{fig:mun_NM}, \ref{fig:mun_pQCD}),
(c) the bounds on the pressure at a given energy density (Figs.~\ref{fig:ep_NM}, \ref{fig:ep_pQCD}).

For the pressure at a given baryon chemical potential, the lattice data only provides an upper bound presented in Figs.~\ref{fig:ppid}.
We showed EoSs characterized by a speed of sound $v_s^2 \lesssim 0.2$ for $\muB \simeq 2000$ MeV are ruled out by this upper bound, as can be seen form Fig.\ref{fig:pcss}. This bound on the $v_s$ could be useful for modeling dense matter realized in heavy-ion collisions \cite{Sorensen:2023zkk}.  

Obtaining bounds on the baryon density at a given baryon chemical potential from the lattice data requires additional input.  We express the pressure as an integral of the baryon density using the method in Ref.~\cite{Komoltsev:2021jzg} so that the pressure inequality can be used.
In the integral, we need to specify either a lower bound $\muL$ or upper bound $\muH$ of the integration interval.
In Sec.~\ref{sec:mun}, we took $\muL$ as the empirical saturation point.
The upper bound on the baryon density plotted in Fig.~\ref{fig:mun_NM} is robust;
the only source of uncertainty in this lower bound is the uncertainty of the lattice calculation.
This result implies that the density jump $\Delta \nB$ in the first-order phase transition, if it exists around the saturation density, cannot be infinitely large, but has to be bounded $\Delta \nB < 10~\nsat$.

In Sec.~\ref{sec:pQCD}, we pinned down the perturbative QCD thermodynamics at $\muH$.
Figure~\ref{fig:mun_pQCD} shows the lower bound on the baryon density in addition to the upper bound.
Aside from the lattice uncertainty, this lower bound is also sensitive to the renormalization scale ambiguity in the running coupling constant;
in this work, we did not include this effect in the lattice bounds.

The bounds on the pressure at a given energy density gives a straightforward interpretation for the stiffness of the equation of state.
The lower and higher pressure at a given energy density correspond to the soft and stiff equations of state, respectively.
In Fig.~\ref{fig:ep_NM}, we plot the lower bound from the lattice data.
This bound only assumes the input from the empirical saturation of nuclear matter and the lattice bound, so it is robust.
Combined with the perturbative QCD thermodynamics at $\muH$, one can also put an upper bound on the energy density-pressure plane as can be seen in Fig.~\ref{fig:ep_pQCD}.
The upper bound is close to the causal extrapolation from the empirical saturation point.
The lower bound is also modified in Fig.~\ref{fig:ep_pQCD} as the equation of state is required to converge on a single point at $\muH$.

Finally, we compared the lattice bound with the integral constraint on the interpolation between the low-density and the high-density reference points imposing the thermodynamic stability and causality.
The results are plotted in Figs.~\ref{fig:mun_integ} and \ref{fig:pmu_integ}.
We found that around the saturation density, the information content of the lattice data is as comparable to that of the perturbative QCD at $\muH = 3000~\text{MeV}$.
These results imply that the synergy between the both QCD-based constraints can further restrict the allowed region of the equation of state.

\begin{acknowledgments}
  We thank
  Tyler Gorda,
  Larry McLerran,
  and Peter Petreczky
  for useful conversations.
  We thank
  Agnieszka Sorensen
  for providing us with the EoS data in Ref.~\cite{Oliinychenko:2022uvy}.
  Y.F.\ is supported by the Japan Society for the Promotion of Science (JSPS) through the Overseas Research Fellowship.
  The work of Y.F.\ and S.R.\ was supported by the INT's U.S. DOE Grant No. DE-FG02-00ER41132.
\end{acknowledgments}

\bibliographystyle{apsrev4-2}
\bibliography{bib_trace}

\begin{thebibliography}{62}%
\makeatletter
\providecommand \@ifxundefined [1]{%
 \@ifx{#1\undefined}
}%
\providecommand \@ifnum [1]{%
 \ifnum #1\expandafter \@firstoftwo
 \else \expandafter \@secondoftwo
 \fi
}%
\providecommand \@ifx [1]{%
 \ifx #1\expandafter \@firstoftwo
 \else \expandafter \@secondoftwo
 \fi
}%
\providecommand \natexlab [1]{#1}%
\providecommand \enquote  [1]{``#1''}%
\providecommand \bibnamefont  [1]{#1}%
\providecommand \bibfnamefont [1]{#1}%
\providecommand \citenamefont [1]{#1}%
\providecommand \href@noop [0]{\@secondoftwo}%
\providecommand \href [0]{\begingroup \@sanitize@url \@href}%
\providecommand \@href[1]{\@@startlink{#1}\@@href}%
\providecommand \@@href[1]{\endgroup#1\@@endlink}%
\providecommand \@sanitize@url [0]{\catcode `\\12\catcode `\$12\catcode
  `\&12\catcode `\#12\catcode `\^12\catcode `\_12\catcode `\%12\relax}%
\providecommand \@@startlink[1]{}%
\providecommand \@@endlink[0]{}%
\providecommand \url  [0]{\begingroup\@sanitize@url \@url }%
\providecommand \@url [1]{\endgroup\@href {#1}{\urlprefix }}%
\providecommand \urlprefix  [0]{URL }%
\providecommand \Eprint [0]{\href }%
\providecommand \doibase [0]{https://doi.org/}%
\providecommand \selectlanguage [0]{\@gobble}%
\providecommand \bibinfo  [0]{\@secondoftwo}%
\providecommand \bibfield  [0]{\@secondoftwo}%
\providecommand \translation [1]{[#1]}%
\providecommand \BibitemOpen [0]{}%
\providecommand \bibitemStop [0]{}%
\providecommand \bibitemNoStop [0]{.\EOS\space}%
\providecommand \EOS [0]{\spacefactor3000\relax}%
\providecommand \BibitemShut  [1]{\csname bibitem#1\endcsname}%
\let\auto@bib@innerbib\@empty
%</preamble>
\bibitem [{\citenamefont {Drischler}\ \emph {et~al.}(2021)\citenamefont
  {Drischler}, \citenamefont {Holt},\ and\ \citenamefont
  {Wellenhofer}}]{Drischler:2021kxf}%
  \BibitemOpen
  \bibfield  {author} {\bibinfo {author} {\bibfnamefont {C.}~\bibnamefont
  {Drischler}}, \bibinfo {author} {\bibfnamefont {J.~W.}\ \bibnamefont
  {Holt}},\ and\ \bibinfo {author} {\bibfnamefont {C.}~\bibnamefont
  {Wellenhofer}},\ }\href {https://doi.org/10.1146/annurev-nucl-102419-041903}
  {\bibfield  {journal} {\bibinfo  {journal} {Ann. Rev. Nucl. Part. Sci.}\
  }\textbf {\bibinfo {volume} {71}},\ \bibinfo {pages} {403} (\bibinfo {year}
  {2021})},\ \Eprint {https://arxiv.org/abs/2101.01709} {arXiv:2101.01709
  [nucl-th]} \BibitemShut {NoStop}%
\bibitem [{\citenamefont {Ghiglieri}\ \emph {et~al.}(2020)\citenamefont
  {Ghiglieri}, \citenamefont {Kurkela}, \citenamefont {Strickland},\ and\
  \citenamefont {Vuorinen}}]{Ghiglieri:2020dpq}%
  \BibitemOpen
  \bibfield  {author} {\bibinfo {author} {\bibfnamefont {J.}~\bibnamefont
  {Ghiglieri}}, \bibinfo {author} {\bibfnamefont {A.}~\bibnamefont {Kurkela}},
  \bibinfo {author} {\bibfnamefont {M.}~\bibnamefont {Strickland}},\ and\
  \bibinfo {author} {\bibfnamefont {A.}~\bibnamefont {Vuorinen}},\ }\href
  {https://doi.org/10.1016/j.physrep.2020.07.004} {\bibfield  {journal}
  {\bibinfo  {journal} {Phys. Rept.}\ }\textbf {\bibinfo {volume} {880}},\
  \bibinfo {pages} {1} (\bibinfo {year} {2020})},\ \Eprint
  {https://arxiv.org/abs/2002.10188} {arXiv:2002.10188 [hep-ph]} \BibitemShut
  {NoStop}%
\bibitem [{\citenamefont {Komoltsev}\ and\ \citenamefont
  {Kurkela}(2022)}]{Komoltsev:2021jzg}%
  \BibitemOpen
  \bibfield  {author} {\bibinfo {author} {\bibfnamefont {O.}~\bibnamefont
  {Komoltsev}}\ and\ \bibinfo {author} {\bibfnamefont {A.}~\bibnamefont
  {Kurkela}},\ }\href {https://doi.org/10.1103/PhysRevLett.128.202701}
  {\bibfield  {journal} {\bibinfo  {journal} {Phys. Rev. Lett.}\ }\textbf
  {\bibinfo {volume} {128}},\ \bibinfo {pages} {202701} (\bibinfo {year}
  {2022})},\ \Eprint {https://arxiv.org/abs/2111.05350} {arXiv:2111.05350
  [nucl-th]} \BibitemShut {NoStop}%
\bibitem [{\citenamefont {{Lindblom}}(1992)}]{1992ApJ...398..569L}%
  \BibitemOpen
  \bibfield  {author} {\bibinfo {author} {\bibfnamefont {L.}~\bibnamefont
  {{Lindblom}}},\ }\href {https://doi.org/10.1086/171882} {\bibfield  {journal}
  {\bibinfo  {journal} {\apj}\ }\textbf {\bibinfo {volume} {398}},\ \bibinfo
  {pages} {569} (\bibinfo {year} {1992})}\BibitemShut {NoStop}%
\bibitem [{\citenamefont {Kurkela}\ \emph {et~al.}(2014)\citenamefont
  {Kurkela}, \citenamefont {Fraga}, \citenamefont {Schaffner-Bielich},\ and\
  \citenamefont {Vuorinen}}]{Kurkela:2014vha}%
  \BibitemOpen
  \bibfield  {author} {\bibinfo {author} {\bibfnamefont {A.}~\bibnamefont
  {Kurkela}}, \bibinfo {author} {\bibfnamefont {E.~S.}\ \bibnamefont {Fraga}},
  \bibinfo {author} {\bibfnamefont {J.}~\bibnamefont {Schaffner-Bielich}},\
  and\ \bibinfo {author} {\bibfnamefont {A.}~\bibnamefont {Vuorinen}},\ }\href
  {https://doi.org/10.1088/0004-637X/789/2/127} {\bibfield  {journal} {\bibinfo
   {journal} {Astrophys. J.}\ }\textbf {\bibinfo {volume} {789}},\ \bibinfo
  {pages} {127} (\bibinfo {year} {2014})},\ \Eprint
  {https://arxiv.org/abs/1402.6618} {arXiv:1402.6618 [astro-ph.HE]}
  \BibitemShut {NoStop}%
\bibitem [{\citenamefont {Annala}\ \emph {et~al.}(2018)\citenamefont {Annala},
  \citenamefont {Gorda}, \citenamefont {Kurkela},\ and\ \citenamefont
  {Vuorinen}}]{Annala:2017llu}%
  \BibitemOpen
  \bibfield  {author} {\bibinfo {author} {\bibfnamefont {E.}~\bibnamefont
  {Annala}}, \bibinfo {author} {\bibfnamefont {T.}~\bibnamefont {Gorda}},
  \bibinfo {author} {\bibfnamefont {A.}~\bibnamefont {Kurkela}},\ and\ \bibinfo
  {author} {\bibfnamefont {A.}~\bibnamefont {Vuorinen}},\ }\href
  {https://doi.org/10.1103/PhysRevLett.120.172703} {\bibfield  {journal}
  {\bibinfo  {journal} {Phys. Rev. Lett.}\ }\textbf {\bibinfo {volume} {120}},\
  \bibinfo {pages} {172703} (\bibinfo {year} {2018})},\ \Eprint
  {https://arxiv.org/abs/1711.02644} {arXiv:1711.02644 [astro-ph.HE]}
  \BibitemShut {NoStop}%
\bibitem [{\citenamefont {Landry}\ \emph {et~al.}(2020)\citenamefont {Landry},
  \citenamefont {Essick},\ and\ \citenamefont
  {Chatziioannou}}]{Landry:2020vaw}%
  \BibitemOpen
  \bibfield  {author} {\bibinfo {author} {\bibfnamefont {P.}~\bibnamefont
  {Landry}}, \bibinfo {author} {\bibfnamefont {R.}~\bibnamefont {Essick}},\
  and\ \bibinfo {author} {\bibfnamefont {K.}~\bibnamefont {Chatziioannou}},\
  }\href {https://doi.org/10.1103/PhysRevD.101.123007} {\bibfield  {journal}
  {\bibinfo  {journal} {Phys. Rev. D}\ }\textbf {\bibinfo {volume} {101}},\
  \bibinfo {pages} {123007} (\bibinfo {year} {2020})},\ \Eprint
  {https://arxiv.org/abs/2003.04880} {arXiv:2003.04880 [astro-ph.HE]}
  \BibitemShut {NoStop}%
\bibitem [{\citenamefont {Annala}\ \emph {et~al.}(2022)\citenamefont {Annala},
  \citenamefont {Gorda}, \citenamefont {Katerini}, \citenamefont {Kurkela},
  \citenamefont {N\"attil\"a}, \citenamefont {Paschalidis},\ and\ \citenamefont
  {Vuorinen}}]{Annala:2021gom}%
  \BibitemOpen
  \bibfield  {author} {\bibinfo {author} {\bibfnamefont {E.}~\bibnamefont
  {Annala}}, \bibinfo {author} {\bibfnamefont {T.}~\bibnamefont {Gorda}},
  \bibinfo {author} {\bibfnamefont {E.}~\bibnamefont {Katerini}}, \bibinfo
  {author} {\bibfnamefont {A.}~\bibnamefont {Kurkela}}, \bibinfo {author}
  {\bibfnamefont {J.}~\bibnamefont {N\"attil\"a}}, \bibinfo {author}
  {\bibfnamefont {V.}~\bibnamefont {Paschalidis}},\ and\ \bibinfo {author}
  {\bibfnamefont {A.}~\bibnamefont {Vuorinen}},\ }\href
  {https://doi.org/10.1103/PhysRevX.12.011058} {\bibfield  {journal} {\bibinfo
  {journal} {Phys. Rev. X}\ }\textbf {\bibinfo {volume} {12}},\ \bibinfo
  {pages} {011058} (\bibinfo {year} {2022})},\ \Eprint
  {https://arxiv.org/abs/2105.05132} {arXiv:2105.05132 [astro-ph.HE]}
  \BibitemShut {NoStop}%
\bibitem [{\citenamefont {Piekarewicz}(2010)}]{Piekarewicz:2009gb}%
  \BibitemOpen
  \bibfield  {author} {\bibinfo {author} {\bibfnamefont {J.}~\bibnamefont
  {Piekarewicz}},\ }\href {https://doi.org/10.1088/0954-3899/37/6/064038}
  {\bibfield  {journal} {\bibinfo  {journal} {J. Phys. G}\ }\textbf {\bibinfo
  {volume} {37}},\ \bibinfo {pages} {064038} (\bibinfo {year} {2010})},\
  \Eprint {https://arxiv.org/abs/0912.5103} {arXiv:0912.5103 [nucl-th]}
  \BibitemShut {NoStop}%
\bibitem [{\citenamefont {Danielewicz}\ \emph {et~al.}(2002)\citenamefont
  {Danielewicz}, \citenamefont {Lacey},\ and\ \citenamefont
  {Lynch}}]{Danielewicz:2002pu}%
  \BibitemOpen
  \bibfield  {author} {\bibinfo {author} {\bibfnamefont {P.}~\bibnamefont
  {Danielewicz}}, \bibinfo {author} {\bibfnamefont {R.}~\bibnamefont {Lacey}},\
  and\ \bibinfo {author} {\bibfnamefont {W.~G.}\ \bibnamefont {Lynch}},\ }\href
  {https://doi.org/10.1126/science.1078070} {\bibfield  {journal} {\bibinfo
  {journal} {Science}\ }\textbf {\bibinfo {volume} {298}},\ \bibinfo {pages}
  {1592} (\bibinfo {year} {2002})},\ \Eprint
  {https://arxiv.org/abs/nucl-th/0208016} {arXiv:nucl-th/0208016} \BibitemShut
  {NoStop}%
\bibitem [{\citenamefont {Oliinychenko}\ \emph {et~al.}(2023)\citenamefont
  {Oliinychenko}, \citenamefont {Sorensen}, \citenamefont {Koch},\ and\
  \citenamefont {McLerran}}]{Oliinychenko:2022uvy}%
  \BibitemOpen
  \bibfield  {author} {\bibinfo {author} {\bibfnamefont {D.}~\bibnamefont
  {Oliinychenko}}, \bibinfo {author} {\bibfnamefont {A.}~\bibnamefont
  {Sorensen}}, \bibinfo {author} {\bibfnamefont {V.}~\bibnamefont {Koch}},\
  and\ \bibinfo {author} {\bibfnamefont {L.}~\bibnamefont {McLerran}},\ }\href
  {https://doi.org/10.1103/PhysRevC.108.034908} {\bibfield  {journal} {\bibinfo
   {journal} {Phys. Rev. C}\ }\textbf {\bibinfo {volume} {108}},\ \bibinfo
  {pages} {034908} (\bibinfo {year} {2023})},\ \Eprint
  {https://arxiv.org/abs/2208.11996} {arXiv:2208.11996 [nucl-th]} \BibitemShut
  {NoStop}%
\bibitem [{\citenamefont {Sorensen}\ \emph {et~al.}(2023)\citenamefont
  {Sorensen} \emph {et~al.}}]{Sorensen:2023zkk}%
  \BibitemOpen
  \bibfield  {author} {\bibinfo {author} {\bibfnamefont {A.}~\bibnamefont
  {Sorensen}} \emph {et~al.},\ }\href@noop {} {\  (\bibinfo {year} {2023})},\
  \Eprint {https://arxiv.org/abs/2301.13253} {arXiv:2301.13253 [nucl-th]}
  \BibitemShut {NoStop}%
\bibitem [{\citenamefont {Son}\ and\ \citenamefont
  {Stephanov}(2001{\natexlab{a}})}]{Son:2000xc}%
  \BibitemOpen
  \bibfield  {author} {\bibinfo {author} {\bibfnamefont {D.~T.}\ \bibnamefont
  {Son}}\ and\ \bibinfo {author} {\bibfnamefont {M.~A.}\ \bibnamefont
  {Stephanov}},\ }\href {https://doi.org/10.1103/PhysRevLett.86.592} {\bibfield
   {journal} {\bibinfo  {journal} {Phys. Rev. Lett.}\ }\textbf {\bibinfo
  {volume} {86}},\ \bibinfo {pages} {592} (\bibinfo {year}
  {2001}{\natexlab{a}})},\ \Eprint {https://arxiv.org/abs/hep-ph/0005225}
  {arXiv:hep-ph/0005225} \BibitemShut {NoStop}%
\bibitem [{\citenamefont {Son}\ and\ \citenamefont
  {Stephanov}(2001{\natexlab{b}})}]{Son:2000by}%
  \BibitemOpen
  \bibfield  {author} {\bibinfo {author} {\bibfnamefont {D.~T.}\ \bibnamefont
  {Son}}\ and\ \bibinfo {author} {\bibfnamefont {M.~A.}\ \bibnamefont
  {Stephanov}},\ }\href {https://doi.org/10.1134/1.1378872} {\bibfield
  {journal} {\bibinfo  {journal} {Phys. Atom. Nucl.}\ }\textbf {\bibinfo
  {volume} {64}},\ \bibinfo {pages} {834} (\bibinfo {year}
  {2001}{\natexlab{b}})},\ \Eprint {https://arxiv.org/abs/hep-ph/0011365}
  {arXiv:hep-ph/0011365} \BibitemShut {NoStop}%
\bibitem [{\citenamefont {Cohen}(2003{\natexlab{a}})}]{Cohen:2003ut}%
  \BibitemOpen
  \bibfield  {author} {\bibinfo {author} {\bibfnamefont {T.~D.}\ \bibnamefont
  {Cohen}},\ }\href {https://doi.org/10.1103/PhysRevLett.91.032002} {\bibfield
  {journal} {\bibinfo  {journal} {Phys. Rev. Lett.}\ }\textbf {\bibinfo
  {volume} {91}},\ \bibinfo {pages} {032002} (\bibinfo {year}
  {2003}{\natexlab{a}})},\ \Eprint {https://arxiv.org/abs/hep-ph/0304024}
  {arXiv:hep-ph/0304024} \BibitemShut {NoStop}%
\bibitem [{\citenamefont {Cohen}(2004)}]{Cohen:2004qp}%
  \BibitemOpen
  \bibfield  {author} {\bibinfo {author} {\bibfnamefont {T.~D.}\ \bibnamefont
  {Cohen}},\ }in\ \href {https://doi.org/10.1142/9789812775344_0009} {\emph
  {\bibinfo {booktitle} {{From Fields to Strings: Circumnavigating Theoretical
  Physics: A Conference in Tribute to Ian Kogan}}}}\ (\bibinfo {year} {2004})\
  pp.\ \bibinfo {pages} {101--120},\ \Eprint
  {https://arxiv.org/abs/hep-ph/0405043} {arXiv:hep-ph/0405043} \BibitemShut
  {NoStop}%
\bibitem [{\citenamefont {Abbott}\ \emph {et~al.}(2023)\citenamefont {Abbott},
  \citenamefont {Detmold}, \citenamefont {Romero-L\'opez}, \citenamefont
  {Davoudi}, \citenamefont {Illa}, \citenamefont {Parre\~no}, \citenamefont
  {Perry}, \citenamefont {Shanahan},\ and\ \citenamefont
  {Wagman}}]{Abbott:2023coj}%
  \BibitemOpen
  \bibfield  {author} {\bibinfo {author} {\bibfnamefont {R.}~\bibnamefont
  {Abbott}}, \bibinfo {author} {\bibfnamefont {W.}~\bibnamefont {Detmold}},
  \bibinfo {author} {\bibfnamefont {F.}~\bibnamefont {Romero-L\'opez}},
  \bibinfo {author} {\bibfnamefont {Z.}~\bibnamefont {Davoudi}}, \bibinfo
  {author} {\bibfnamefont {M.}~\bibnamefont {Illa}}, \bibinfo {author}
  {\bibfnamefont {A.}~\bibnamefont {Parre\~no}}, \bibinfo {author}
  {\bibfnamefont {R.~J.}\ \bibnamefont {Perry}}, \bibinfo {author}
  {\bibfnamefont {P.~E.}\ \bibnamefont {Shanahan}},\ and\ \bibinfo {author}
  {\bibfnamefont {M.~L.}\ \bibnamefont {Wagman}},\ }\href@noop {} {\  (\bibinfo
  {year} {2023})},\ \Eprint {https://arxiv.org/abs/2307.15014}
  {arXiv:2307.15014 [hep-lat]} \BibitemShut {NoStop}%
\bibitem [{\citenamefont {Weingarten}(1983)}]{Weingarten:1983uj}%
  \BibitemOpen
  \bibfield  {author} {\bibinfo {author} {\bibfnamefont {D.}~\bibnamefont
  {Weingarten}},\ }\href {https://doi.org/10.1103/PhysRevLett.51.1830}
  {\bibfield  {journal} {\bibinfo  {journal} {Phys. Rev. Lett.}\ }\textbf
  {\bibinfo {volume} {51}},\ \bibinfo {pages} {1830} (\bibinfo {year}
  {1983})}\BibitemShut {NoStop}%
\bibitem [{\citenamefont {Witten}(1983)}]{Witten:1983ut}%
  \BibitemOpen
  \bibfield  {author} {\bibinfo {author} {\bibfnamefont {E.}~\bibnamefont
  {Witten}},\ }\href {https://doi.org/10.1103/PhysRevLett.51.2351} {\bibfield
  {journal} {\bibinfo  {journal} {Phys. Rev. Lett.}\ }\textbf {\bibinfo
  {volume} {51}},\ \bibinfo {pages} {2351} (\bibinfo {year}
  {1983})}\BibitemShut {NoStop}%
\bibitem [{\citenamefont {Vafa}\ and\ \citenamefont
  {Witten}(1984)}]{Vafa:1983tf}%
  \BibitemOpen
  \bibfield  {author} {\bibinfo {author} {\bibfnamefont {C.}~\bibnamefont
  {Vafa}}\ and\ \bibinfo {author} {\bibfnamefont {E.}~\bibnamefont {Witten}},\
  }\href {https://doi.org/10.1016/0550-3213(84)90230-X} {\bibfield  {journal}
  {\bibinfo  {journal} {Nucl. Phys. B}\ }\textbf {\bibinfo {volume} {234}},\
  \bibinfo {pages} {173} (\bibinfo {year} {1984})}\BibitemShut {NoStop}%
\bibitem [{\citenamefont {Nussinov}(1983)}]{Nussinov:1983vh}%
  \BibitemOpen
  \bibfield  {author} {\bibinfo {author} {\bibfnamefont {S.}~\bibnamefont
  {Nussinov}},\ }\href {https://doi.org/10.1103/PhysRevLett.51.2081} {\bibfield
   {journal} {\bibinfo  {journal} {Phys. Rev. Lett.}\ }\textbf {\bibinfo
  {volume} {51}},\ \bibinfo {pages} {2081} (\bibinfo {year}
  {1983})}\BibitemShut {NoStop}%
\bibitem [{\citenamefont {Nussinov}\ and\ \citenamefont
  {Lampert}(2002)}]{Nussinov:1999sx}%
  \BibitemOpen
  \bibfield  {author} {\bibinfo {author} {\bibfnamefont {S.}~\bibnamefont
  {Nussinov}}\ and\ \bibinfo {author} {\bibfnamefont {M.~A.}\ \bibnamefont
  {Lampert}},\ }\href {https://doi.org/10.1016/S0370-1573(01)00091-6}
  {\bibfield  {journal} {\bibinfo  {journal} {Phys. Rept.}\ }\textbf {\bibinfo
  {volume} {362}},\ \bibinfo {pages} {193} (\bibinfo {year} {2002})},\ \Eprint
  {https://arxiv.org/abs/hep-ph/9911532} {arXiv:hep-ph/9911532} \BibitemShut
  {NoStop}%
\bibitem [{\citenamefont {Alford}\ \emph {et~al.}(1999)\citenamefont {Alford},
  \citenamefont {Kapustin},\ and\ \citenamefont {Wilczek}}]{Alford:1998sd}%
  \BibitemOpen
  \bibfield  {author} {\bibinfo {author} {\bibfnamefont {M.~G.}\ \bibnamefont
  {Alford}}, \bibinfo {author} {\bibfnamefont {A.}~\bibnamefont {Kapustin}},\
  and\ \bibinfo {author} {\bibfnamefont {F.}~\bibnamefont {Wilczek}},\ }\href
  {https://doi.org/10.1103/PhysRevD.59.054502} {\bibfield  {journal} {\bibinfo
  {journal} {Phys. Rev. D}\ }\textbf {\bibinfo {volume} {59}},\ \bibinfo
  {pages} {054502} (\bibinfo {year} {1999})},\ \Eprint
  {https://arxiv.org/abs/hep-lat/9807039} {arXiv:hep-lat/9807039} \BibitemShut
  {NoStop}%
\bibitem [{\citenamefont {Moore}\ and\ \citenamefont
  {Gorda}(2023)}]{Moore:2023glb}%
  \BibitemOpen
  \bibfield  {author} {\bibinfo {author} {\bibfnamefont {G.~D.}\ \bibnamefont
  {Moore}}\ and\ \bibinfo {author} {\bibfnamefont {T.}~\bibnamefont {Gorda}},\
  }\href@noop {} {\  (\bibinfo {year} {2023})},\ \Eprint
  {https://arxiv.org/abs/2309.15149} {arXiv:2309.15149 [nucl-th]} \BibitemShut
  {NoStop}%
\bibitem [{\citenamefont {Hidaka}\ and\ \citenamefont
  {Yamamoto}(2012)}]{Hidaka:2011jj}%
  \BibitemOpen
  \bibfield  {author} {\bibinfo {author} {\bibfnamefont {Y.}~\bibnamefont
  {Hidaka}}\ and\ \bibinfo {author} {\bibfnamefont {N.}~\bibnamefont
  {Yamamoto}},\ }\href {https://doi.org/10.1103/PhysRevLett.108.121601}
  {\bibfield  {journal} {\bibinfo  {journal} {Phys. Rev. Lett.}\ }\textbf
  {\bibinfo {volume} {108}},\ \bibinfo {pages} {121601} (\bibinfo {year}
  {2012})},\ \Eprint {https://arxiv.org/abs/1110.3044} {arXiv:1110.3044
  [hep-ph]} \BibitemShut {NoStop}%
\bibitem [{\citenamefont {Kogut}\ and\ \citenamefont
  {Sinclair}(2002{\natexlab{a}})}]{Kogut:2002tm}%
  \BibitemOpen
  \bibfield  {author} {\bibinfo {author} {\bibfnamefont {J.~B.}\ \bibnamefont
  {Kogut}}\ and\ \bibinfo {author} {\bibfnamefont {D.~K.}\ \bibnamefont
  {Sinclair}},\ }\href {https://doi.org/10.1103/PhysRevD.66.014508} {\bibfield
  {journal} {\bibinfo  {journal} {Phys. Rev. D}\ }\textbf {\bibinfo {volume}
  {66}},\ \bibinfo {pages} {014508} (\bibinfo {year} {2002}{\natexlab{a}})},\
  \Eprint {https://arxiv.org/abs/hep-lat/0201017} {arXiv:hep-lat/0201017}
  \BibitemShut {NoStop}%
\bibitem [{\citenamefont {Kogut}\ and\ \citenamefont
  {Sinclair}(2002{\natexlab{b}})}]{Kogut:2002zg}%
  \BibitemOpen
  \bibfield  {author} {\bibinfo {author} {\bibfnamefont {J.~B.}\ \bibnamefont
  {Kogut}}\ and\ \bibinfo {author} {\bibfnamefont {D.~K.}\ \bibnamefont
  {Sinclair}},\ }\href {https://doi.org/10.1103/PhysRevD.66.034505} {\bibfield
  {journal} {\bibinfo  {journal} {Phys. Rev. D}\ }\textbf {\bibinfo {volume}
  {66}},\ \bibinfo {pages} {034505} (\bibinfo {year} {2002}{\natexlab{b}})},\
  \Eprint {https://arxiv.org/abs/hep-lat/0202028} {arXiv:hep-lat/0202028}
  \BibitemShut {NoStop}%
\bibitem [{\citenamefont {Kogut}\ and\ \citenamefont
  {Sinclair}(2004)}]{Kogut:2004zg}%
  \BibitemOpen
  \bibfield  {author} {\bibinfo {author} {\bibfnamefont {J.~B.}\ \bibnamefont
  {Kogut}}\ and\ \bibinfo {author} {\bibfnamefont {D.~K.}\ \bibnamefont
  {Sinclair}},\ }\href {https://doi.org/10.1103/PhysRevD.70.094501} {\bibfield
  {journal} {\bibinfo  {journal} {Phys. Rev. D}\ }\textbf {\bibinfo {volume}
  {70}},\ \bibinfo {pages} {094501} (\bibinfo {year} {2004})},\ \Eprint
  {https://arxiv.org/abs/hep-lat/0407027} {arXiv:hep-lat/0407027} \BibitemShut
  {NoStop}%
\bibitem [{\citenamefont {de~Forcrand}\ \emph {et~al.}(2007)\citenamefont
  {de~Forcrand}, \citenamefont {Stephanov},\ and\ \citenamefont
  {Wenger}}]{deForcrand:2007uz}%
  \BibitemOpen
  \bibfield  {author} {\bibinfo {author} {\bibfnamefont {P.}~\bibnamefont
  {de~Forcrand}}, \bibinfo {author} {\bibfnamefont {M.~A.}\ \bibnamefont
  {Stephanov}},\ and\ \bibinfo {author} {\bibfnamefont {U.}~\bibnamefont
  {Wenger}},\ }\href@noop {} {\bibfield  {journal} {\bibinfo  {journal} {PoS}\
  }\textbf {\bibinfo {volume} {LATTICE2007}},\ \bibinfo {pages} {237} (\bibinfo
  {year} {2007})},\ \Eprint {https://arxiv.org/abs/0711.0023} {arXiv:0711.0023
  [hep-lat]} \BibitemShut {NoStop}%
\bibitem [{\citenamefont {Detmold}\ \emph {et~al.}(2008)\citenamefont
  {Detmold}, \citenamefont {Savage}, \citenamefont {Torok}, \citenamefont
  {Beane}, \citenamefont {Luu}, \citenamefont {Orginos},\ and\ \citenamefont
  {Parreno}}]{Detmold:2008fn}%
  \BibitemOpen
  \bibfield  {author} {\bibinfo {author} {\bibfnamefont {W.}~\bibnamefont
  {Detmold}}, \bibinfo {author} {\bibfnamefont {M.~J.}\ \bibnamefont {Savage}},
  \bibinfo {author} {\bibfnamefont {A.}~\bibnamefont {Torok}}, \bibinfo
  {author} {\bibfnamefont {S.~R.}\ \bibnamefont {Beane}}, \bibinfo {author}
  {\bibfnamefont {T.~C.}\ \bibnamefont {Luu}}, \bibinfo {author} {\bibfnamefont
  {K.}~\bibnamefont {Orginos}},\ and\ \bibinfo {author} {\bibfnamefont
  {A.}~\bibnamefont {Parreno}},\ }\href
  {https://doi.org/10.1103/PhysRevD.78.014507} {\bibfield  {journal} {\bibinfo
  {journal} {Phys. Rev. D}\ }\textbf {\bibinfo {volume} {78}},\ \bibinfo
  {pages} {014507} (\bibinfo {year} {2008})},\ \Eprint
  {https://arxiv.org/abs/0803.2728} {arXiv:0803.2728 [hep-lat]} \BibitemShut
  {NoStop}%
\bibitem [{\citenamefont {Cea}\ \emph {et~al.}(2012)\citenamefont {Cea},
  \citenamefont {Cosmai}, \citenamefont {D'Elia}, \citenamefont {Papa},\ and\
  \citenamefont {Sanfilippo}}]{Cea:2012ev}%
  \BibitemOpen
  \bibfield  {author} {\bibinfo {author} {\bibfnamefont {P.}~\bibnamefont
  {Cea}}, \bibinfo {author} {\bibfnamefont {L.}~\bibnamefont {Cosmai}},
  \bibinfo {author} {\bibfnamefont {M.}~\bibnamefont {D'Elia}}, \bibinfo
  {author} {\bibfnamefont {A.}~\bibnamefont {Papa}},\ and\ \bibinfo {author}
  {\bibfnamefont {F.}~\bibnamefont {Sanfilippo}},\ }\href
  {https://doi.org/10.1103/PhysRevD.85.094512} {\bibfield  {journal} {\bibinfo
  {journal} {Phys. Rev. D}\ }\textbf {\bibinfo {volume} {85}},\ \bibinfo
  {pages} {094512} (\bibinfo {year} {2012})},\ \Eprint
  {https://arxiv.org/abs/1202.5700} {arXiv:1202.5700 [hep-lat]} \BibitemShut
  {NoStop}%
\bibitem [{\citenamefont {Detmold}\ \emph {et~al.}(2012)\citenamefont
  {Detmold}, \citenamefont {Orginos},\ and\ \citenamefont
  {Shi}}]{Detmold:2012wc}%
  \BibitemOpen
  \bibfield  {author} {\bibinfo {author} {\bibfnamefont {W.}~\bibnamefont
  {Detmold}}, \bibinfo {author} {\bibfnamefont {K.}~\bibnamefont {Orginos}},\
  and\ \bibinfo {author} {\bibfnamefont {Z.}~\bibnamefont {Shi}},\ }\href
  {https://doi.org/10.1103/PhysRevD.86.054507} {\bibfield  {journal} {\bibinfo
  {journal} {Phys. Rev. D}\ }\textbf {\bibinfo {volume} {86}},\ \bibinfo
  {pages} {054507} (\bibinfo {year} {2012})},\ \Eprint
  {https://arxiv.org/abs/1205.4224} {arXiv:1205.4224 [hep-lat]} \BibitemShut
  {NoStop}%
\bibitem [{\citenamefont {Detmold}\ \emph {et~al.}(2013)\citenamefont
  {Detmold}, \citenamefont {Meinel},\ and\ \citenamefont
  {Shi}}]{Detmold:2012pi}%
  \BibitemOpen
  \bibfield  {author} {\bibinfo {author} {\bibfnamefont {W.}~\bibnamefont
  {Detmold}}, \bibinfo {author} {\bibfnamefont {S.}~\bibnamefont {Meinel}},\
  and\ \bibinfo {author} {\bibfnamefont {Z.}~\bibnamefont {Shi}},\ }\href
  {https://doi.org/10.1103/PhysRevD.87.094504} {\bibfield  {journal} {\bibinfo
  {journal} {Phys. Rev. D}\ }\textbf {\bibinfo {volume} {87}},\ \bibinfo
  {pages} {094504} (\bibinfo {year} {2013})},\ \Eprint
  {https://arxiv.org/abs/1211.3156} {arXiv:1211.3156 [hep-lat]} \BibitemShut
  {NoStop}%
\bibitem [{\citenamefont {Endr\"odi}(2014)}]{Endrodi:2014lja}%
  \BibitemOpen
  \bibfield  {author} {\bibinfo {author} {\bibfnamefont {G.}~\bibnamefont
  {Endr\"odi}},\ }\href {https://doi.org/10.1103/PhysRevD.90.094501} {\bibfield
   {journal} {\bibinfo  {journal} {Phys. Rev. D}\ }\textbf {\bibinfo {volume}
  {90}},\ \bibinfo {pages} {094501} (\bibinfo {year} {2014})},\ \Eprint
  {https://arxiv.org/abs/1407.1216} {arXiv:1407.1216 [hep-lat]} \BibitemShut
  {NoStop}%
\bibitem [{\citenamefont {Brandt}\ \emph {et~al.}(2018)\citenamefont {Brandt},
  \citenamefont {Endrodi},\ and\ \citenamefont
  {Schmalzbauer}}]{Brandt:2017oyy}%
  \BibitemOpen
  \bibfield  {author} {\bibinfo {author} {\bibfnamefont {B.~B.}\ \bibnamefont
  {Brandt}}, \bibinfo {author} {\bibfnamefont {G.}~\bibnamefont {Endrodi}},\
  and\ \bibinfo {author} {\bibfnamefont {S.}~\bibnamefont {Schmalzbauer}},\
  }\href {https://doi.org/10.1103/PhysRevD.97.054514} {\bibfield  {journal}
  {\bibinfo  {journal} {Phys. Rev. D}\ }\textbf {\bibinfo {volume} {97}},\
  \bibinfo {pages} {054514} (\bibinfo {year} {2018})},\ \Eprint
  {https://arxiv.org/abs/1712.08190} {arXiv:1712.08190 [hep-lat]} \BibitemShut
  {NoStop}%
\bibitem [{\citenamefont {Brandt}\ and\ \citenamefont
  {Endrodi}(2019)}]{Brandt:2018omg}%
  \BibitemOpen
  \bibfield  {author} {\bibinfo {author} {\bibfnamefont {B.~B.}\ \bibnamefont
  {Brandt}}\ and\ \bibinfo {author} {\bibfnamefont {G.}~\bibnamefont
  {Endrodi}},\ }\href {https://doi.org/10.1103/PhysRevD.99.014518} {\bibfield
  {journal} {\bibinfo  {journal} {Phys. Rev. D}\ }\textbf {\bibinfo {volume}
  {99}},\ \bibinfo {pages} {014518} (\bibinfo {year} {2019})},\ \Eprint
  {https://arxiv.org/abs/1810.11045} {arXiv:1810.11045 [hep-lat]} \BibitemShut
  {NoStop}%
\bibitem [{\citenamefont {Brandt}\ \emph
  {et~al.}(2023{\natexlab{a}})\citenamefont {Brandt}, \citenamefont {Cuteri},\
  and\ \citenamefont {Endrodi}}]{Brandt:2022hwy}%
  \BibitemOpen
  \bibfield  {author} {\bibinfo {author} {\bibfnamefont {B.~B.}\ \bibnamefont
  {Brandt}}, \bibinfo {author} {\bibfnamefont {F.}~\bibnamefont {Cuteri}},\
  and\ \bibinfo {author} {\bibfnamefont {G.}~\bibnamefont {Endrodi}},\ }\href
  {https://doi.org/10.1007/JHEP07(2023)055} {\bibfield  {journal} {\bibinfo
  {journal} {JHEP}\ }\textbf {\bibinfo {volume} {07}},\ \bibinfo {pages}
  {055}},\ \Eprint {https://arxiv.org/abs/2212.14016} {arXiv:2212.14016
  [hep-lat]} \BibitemShut {NoStop}%
\bibitem [{\citenamefont {Brandt}\ \emph
  {et~al.}(2023{\natexlab{b}})\citenamefont {Brandt}, \citenamefont
  {Chelnokov}, \citenamefont {Cuteri},\ and\ \citenamefont
  {Endr\H{o}di}}]{Brandt:2023kev}%
  \BibitemOpen
  \bibfield  {author} {\bibinfo {author} {\bibfnamefont {B.~B.}\ \bibnamefont
  {Brandt}}, \bibinfo {author} {\bibfnamefont {V.}~\bibnamefont {Chelnokov}},
  \bibinfo {author} {\bibfnamefont {F.}~\bibnamefont {Cuteri}},\ and\ \bibinfo
  {author} {\bibfnamefont {G.}~\bibnamefont {Endr\H{o}di}},\ }\href
  {https://doi.org/10.22323/1.430.0146} {\bibfield  {journal} {\bibinfo
  {journal} {PoS}\ }\textbf {\bibinfo {volume} {LATTICE2022}},\ \bibinfo
  {pages} {146} (\bibinfo {year} {2023}{\natexlab{b}})},\ \Eprint
  {https://arxiv.org/abs/2301.08607} {arXiv:2301.08607 [hep-lat]} \BibitemShut
  {NoStop}%
\bibitem [{\citenamefont {Hands}\ \emph {et~al.}(2006)\citenamefont {Hands},
  \citenamefont {Kim},\ and\ \citenamefont {Skullerud}}]{Hands:2006ve}%
  \BibitemOpen
  \bibfield  {author} {\bibinfo {author} {\bibfnamefont {S.}~\bibnamefont
  {Hands}}, \bibinfo {author} {\bibfnamefont {S.}~\bibnamefont {Kim}},\ and\
  \bibinfo {author} {\bibfnamefont {J.-I.}\ \bibnamefont {Skullerud}},\ }\href
  {https://doi.org/10.1140/epjc/s2006-02621-8} {\bibfield  {journal} {\bibinfo
  {journal} {Eur. Phys. J. C}\ }\textbf {\bibinfo {volume} {48}},\ \bibinfo
  {pages} {193} (\bibinfo {year} {2006})},\ \Eprint
  {https://arxiv.org/abs/hep-lat/0604004} {arXiv:hep-lat/0604004} \BibitemShut
  {NoStop}%
\bibitem [{\citenamefont {Cotter}\ \emph {et~al.}(2013)\citenamefont {Cotter},
  \citenamefont {Giudice}, \citenamefont {Hands},\ and\ \citenamefont
  {Skullerud}}]{Cotter:2012mb}%
  \BibitemOpen
  \bibfield  {author} {\bibinfo {author} {\bibfnamefont {S.}~\bibnamefont
  {Cotter}}, \bibinfo {author} {\bibfnamefont {P.}~\bibnamefont {Giudice}},
  \bibinfo {author} {\bibfnamefont {S.}~\bibnamefont {Hands}},\ and\ \bibinfo
  {author} {\bibfnamefont {J.-I.}\ \bibnamefont {Skullerud}},\ }\href
  {https://doi.org/10.1103/PhysRevD.87.034507} {\bibfield  {journal} {\bibinfo
  {journal} {Phys. Rev. D}\ }\textbf {\bibinfo {volume} {87}},\ \bibinfo
  {pages} {034507} (\bibinfo {year} {2013})},\ \Eprint
  {https://arxiv.org/abs/1210.4496} {arXiv:1210.4496 [hep-lat]} \BibitemShut
  {NoStop}%
\bibitem [{\citenamefont {Boz}\ \emph {et~al.}(2020)\citenamefont {Boz},
  \citenamefont {Giudice}, \citenamefont {Hands},\ and\ \citenamefont
  {Skullerud}}]{Boz:2019enj}%
  \BibitemOpen
  \bibfield  {author} {\bibinfo {author} {\bibfnamefont {T.}~\bibnamefont
  {Boz}}, \bibinfo {author} {\bibfnamefont {P.}~\bibnamefont {Giudice}},
  \bibinfo {author} {\bibfnamefont {S.}~\bibnamefont {Hands}},\ and\ \bibinfo
  {author} {\bibfnamefont {J.-I.}\ \bibnamefont {Skullerud}},\ }\href
  {https://doi.org/10.1103/PhysRevD.101.074506} {\bibfield  {journal} {\bibinfo
   {journal} {Phys. Rev. D}\ }\textbf {\bibinfo {volume} {101}},\ \bibinfo
  {pages} {074506} (\bibinfo {year} {2020})},\ \Eprint
  {https://arxiv.org/abs/1912.10975} {arXiv:1912.10975 [hep-lat]} \BibitemShut
  {NoStop}%
\bibitem [{\citenamefont {Begun}\ \emph {et~al.}(2022)\citenamefont {Begun},
  \citenamefont {Bornyakov}, \citenamefont {Goy}, \citenamefont {Nakamura},\
  and\ \citenamefont {Rogalyov}}]{Begun:2022bxj}%
  \BibitemOpen
  \bibfield  {author} {\bibinfo {author} {\bibfnamefont {A.}~\bibnamefont
  {Begun}}, \bibinfo {author} {\bibfnamefont {V.~G.}\ \bibnamefont
  {Bornyakov}}, \bibinfo {author} {\bibfnamefont {V.~A.}\ \bibnamefont {Goy}},
  \bibinfo {author} {\bibfnamefont {A.}~\bibnamefont {Nakamura}},\ and\
  \bibinfo {author} {\bibfnamefont {R.~N.}\ \bibnamefont {Rogalyov}},\ }\href
  {https://doi.org/10.1103/PhysRevD.105.114505} {\bibfield  {journal} {\bibinfo
   {journal} {Phys. Rev. D}\ }\textbf {\bibinfo {volume} {105}},\ \bibinfo
  {pages} {114505} (\bibinfo {year} {2022})},\ \Eprint
  {https://arxiv.org/abs/2203.04909} {arXiv:2203.04909 [hep-lat]} \BibitemShut
  {NoStop}%
\bibitem [{\citenamefont {Iida}\ and\ \citenamefont
  {Itou}(2022)}]{Iida:2022hyy}%
  \BibitemOpen
  \bibfield  {author} {\bibinfo {author} {\bibfnamefont {K.}~\bibnamefont
  {Iida}}\ and\ \bibinfo {author} {\bibfnamefont {E.}~\bibnamefont {Itou}},\
  }\href {https://doi.org/10.1093/ptep/ptac137} {\bibfield  {journal} {\bibinfo
   {journal} {PTEP}\ }\textbf {\bibinfo {volume} {2022}},\ \bibinfo {pages}
  {111B01} (\bibinfo {year} {2022})},\ \Eprint
  {https://arxiv.org/abs/2207.01253} {arXiv:2207.01253 [hep-ph]} \BibitemShut
  {NoStop}%
\bibitem [{\citenamefont {Stephanov}(2004)}]{Stephanov:2004wx}%
  \BibitemOpen
  \bibfield  {author} {\bibinfo {author} {\bibfnamefont {M.~A.}\ \bibnamefont
  {Stephanov}},\ }\href {https://doi.org/10.1142/S0217751X05027965} {\bibfield
  {journal} {\bibinfo  {journal} {Prog. Theor. Phys. Suppl.}\ }\textbf
  {\bibinfo {volume} {153}},\ \bibinfo {pages} {139} (\bibinfo {year}
  {2004})},\ \Eprint {https://arxiv.org/abs/hep-ph/0402115}
  {arXiv:hep-ph/0402115} \BibitemShut {NoStop}%
\bibitem [{\citenamefont {Bzdak}\ \emph {et~al.}(2020)\citenamefont {Bzdak},
  \citenamefont {Esumi}, \citenamefont {Koch}, \citenamefont {Liao},
  \citenamefont {Stephanov},\ and\ \citenamefont {Xu}}]{Bzdak:2019pkr}%
  \BibitemOpen
  \bibfield  {author} {\bibinfo {author} {\bibfnamefont {A.}~\bibnamefont
  {Bzdak}}, \bibinfo {author} {\bibfnamefont {S.}~\bibnamefont {Esumi}},
  \bibinfo {author} {\bibfnamefont {V.}~\bibnamefont {Koch}}, \bibinfo {author}
  {\bibfnamefont {J.}~\bibnamefont {Liao}}, \bibinfo {author} {\bibfnamefont
  {M.}~\bibnamefont {Stephanov}},\ and\ \bibinfo {author} {\bibfnamefont
  {N.}~\bibnamefont {Xu}},\ }\href
  {https://doi.org/10.1016/j.physrep.2020.01.005} {\bibfield  {journal}
  {\bibinfo  {journal} {Phys. Rept.}\ }\textbf {\bibinfo {volume} {853}},\
  \bibinfo {pages} {1} (\bibinfo {year} {2020})},\ \Eprint
  {https://arxiv.org/abs/1906.00936} {arXiv:1906.00936 [nucl-th]} \BibitemShut
  {NoStop}%
\bibitem [{\citenamefont {Cohen}(2003{\natexlab{b}})}]{Cohen:2003kd}%
  \BibitemOpen
  \bibfield  {author} {\bibinfo {author} {\bibfnamefont {T.~D.}\ \bibnamefont
  {Cohen}},\ }\href {https://doi.org/10.1103/PhysRevLett.91.222001} {\bibfield
  {journal} {\bibinfo  {journal} {Phys. Rev. Lett.}\ }\textbf {\bibinfo
  {volume} {91}},\ \bibinfo {pages} {222001} (\bibinfo {year}
  {2003}{\natexlab{b}})},\ \Eprint {https://arxiv.org/abs/hep-ph/0307089}
  {arXiv:hep-ph/0307089} \BibitemShut {NoStop}%
\bibitem [{\citenamefont {de~Forcrand}\ \emph {et~al.}(2003)\citenamefont
  {de~Forcrand}, \citenamefont {Kim},\ and\ \citenamefont
  {Takaishi}}]{deForcrand:2002pa}%
  \BibitemOpen
  \bibfield  {author} {\bibinfo {author} {\bibfnamefont {P.}~\bibnamefont
  {de~Forcrand}}, \bibinfo {author} {\bibfnamefont {S.}~\bibnamefont {Kim}},\
  and\ \bibinfo {author} {\bibfnamefont {T.}~\bibnamefont {Takaishi}},\ }\href
  {https://doi.org/10.1016/S0920-5632(03)80451-6} {\bibfield  {journal}
  {\bibinfo  {journal} {Nucl. Phys. B Proc. Suppl.}\ }\textbf {\bibinfo
  {volume} {119}},\ \bibinfo {pages} {541} (\bibinfo {year} {2003})},\ \Eprint
  {https://arxiv.org/abs/hep-lat/0209126} {arXiv:hep-lat/0209126} \BibitemShut
  {NoStop}%
\bibitem [{\citenamefont {de~Forcrand}(2009)}]{deForcrand:2009zkb}%
  \BibitemOpen
  \bibfield  {author} {\bibinfo {author} {\bibfnamefont {P.}~\bibnamefont
  {de~Forcrand}},\ }\href {https://doi.org/10.22323/1.091.0010} {\bibfield
  {journal} {\bibinfo  {journal} {PoS}\ }\textbf {\bibinfo {volume}
  {LAT2009}},\ \bibinfo {pages} {010} (\bibinfo {year} {2009})},\ \Eprint
  {https://arxiv.org/abs/1005.0539} {arXiv:1005.0539 [hep-lat]} \BibitemShut
  {NoStop}%
\bibitem [{\citenamefont {Hsu}\ and\ \citenamefont {Reeb}(2010)}]{Hsu:2010zza}%
  \BibitemOpen
  \bibfield  {author} {\bibinfo {author} {\bibfnamefont {S.~D.~H.}\
  \bibnamefont {Hsu}}\ and\ \bibinfo {author} {\bibfnamefont {D.}~\bibnamefont
  {Reeb}},\ }\href {https://doi.org/10.1142/S0217751X10047968} {\bibfield
  {journal} {\bibinfo  {journal} {Int. J. Mod. Phys. A}\ }\textbf {\bibinfo
  {volume} {25}},\ \bibinfo {pages} {53} (\bibinfo {year} {2010})}\BibitemShut
  {NoStop}%
\bibitem [{\citenamefont {Giordano}\ \emph {et~al.}(2020)\citenamefont
  {Giordano}, \citenamefont {Kapas}, \citenamefont {Katz}, \citenamefont
  {Nogradi},\ and\ \citenamefont {Pasztor}}]{Giordano:2020roi}%
  \BibitemOpen
  \bibfield  {author} {\bibinfo {author} {\bibfnamefont {M.}~\bibnamefont
  {Giordano}}, \bibinfo {author} {\bibfnamefont {K.}~\bibnamefont {Kapas}},
  \bibinfo {author} {\bibfnamefont {S.~D.}\ \bibnamefont {Katz}}, \bibinfo
  {author} {\bibfnamefont {D.}~\bibnamefont {Nogradi}},\ and\ \bibinfo {author}
  {\bibfnamefont {A.}~\bibnamefont {Pasztor}},\ }\href
  {https://doi.org/10.1007/JHEP05(2020)088} {\bibfield  {journal} {\bibinfo
  {journal} {JHEP}\ }\textbf {\bibinfo {volume} {05}},\ \bibinfo {pages}
  {088}},\ \Eprint {https://arxiv.org/abs/2004.10800} {arXiv:2004.10800
  [hep-lat]} \BibitemShut {NoStop}%
\bibitem [{\citenamefont {Borsanyi}\ \emph {et~al.}(2022)\citenamefont
  {Borsanyi}, \citenamefont {Fodor}, \citenamefont {Kapas}, \citenamefont
  {Giordano}, \citenamefont {Nogradi}, \citenamefont {Pasztor}, \citenamefont
  {Katz},\ and\ \citenamefont {Wong}}]{Borsanyi:2021hgr}%
  \BibitemOpen
  \bibfield  {author} {\bibinfo {author} {\bibfnamefont {S.}~\bibnamefont
  {Borsanyi}}, \bibinfo {author} {\bibfnamefont {Z.}~\bibnamefont {Fodor}},
  \bibinfo {author} {\bibfnamefont {K.}~\bibnamefont {Kapas}}, \bibinfo
  {author} {\bibfnamefont {M.}~\bibnamefont {Giordano}}, \bibinfo {author}
  {\bibfnamefont {D.}~\bibnamefont {Nogradi}}, \bibinfo {author} {\bibfnamefont
  {A.}~\bibnamefont {Pasztor}}, \bibinfo {author} {\bibfnamefont {S.~D.}\
  \bibnamefont {Katz}},\ and\ \bibinfo {author} {\bibfnamefont {C.~H.}\
  \bibnamefont {Wong}},\ }\href {https://doi.org/10.22323/1.396.0128}
  {\bibfield  {journal} {\bibinfo  {journal} {PoS}\ }\textbf {\bibinfo {volume}
  {LATTICE2021}},\ \bibinfo {pages} {128} (\bibinfo {year} {2022})},\ \Eprint
  {https://arxiv.org/abs/2112.02134} {arXiv:2112.02134 [hep-lat]} \BibitemShut
  {NoStop}%
\bibitem [{\citenamefont {Lee}(2005)}]{Lee:2004hc}%
  \BibitemOpen
  \bibfield  {author} {\bibinfo {author} {\bibfnamefont {D.}~\bibnamefont
  {Lee}},\ }\href {https://doi.org/10.1103/PhysRevC.71.044001} {\bibfield
  {journal} {\bibinfo  {journal} {Phys. Rev. C}\ }\textbf {\bibinfo {volume}
  {71}},\ \bibinfo {pages} {044001} (\bibinfo {year} {2005})},\ \Eprint
  {https://arxiv.org/abs/nucl-th/0407101} {arXiv:nucl-th/0407101} \BibitemShut
  {NoStop}%
\bibitem [{\citenamefont {Freedman}\ and\ \citenamefont
  {McLerran}(1977{\natexlab{a}})}]{Freedman:1976xs}%
  \BibitemOpen
  \bibfield  {author} {\bibinfo {author} {\bibfnamefont {B.~A.}\ \bibnamefont
  {Freedman}}\ and\ \bibinfo {author} {\bibfnamefont {L.~D.}\ \bibnamefont
  {McLerran}},\ }\href {https://doi.org/10.1103/PhysRevD.16.1130} {\bibfield
  {journal} {\bibinfo  {journal} {Phys. Rev. D}\ }\textbf {\bibinfo {volume}
  {16}},\ \bibinfo {pages} {1130} (\bibinfo {year}
  {1977}{\natexlab{a}})}\BibitemShut {NoStop}%
\bibitem [{\citenamefont {Freedman}\ and\ \citenamefont
  {McLerran}(1977{\natexlab{b}})}]{Freedman:1976dm}%
  \BibitemOpen
  \bibfield  {author} {\bibinfo {author} {\bibfnamefont {B.~A.}\ \bibnamefont
  {Freedman}}\ and\ \bibinfo {author} {\bibfnamefont {L.~D.}\ \bibnamefont
  {McLerran}},\ }\href {https://doi.org/10.1103/PhysRevD.16.1147} {\bibfield
  {journal} {\bibinfo  {journal} {Phys. Rev. D}\ }\textbf {\bibinfo {volume}
  {16}},\ \bibinfo {pages} {1147} (\bibinfo {year}
  {1977}{\natexlab{b}})}\BibitemShut {NoStop}%
\bibitem [{\citenamefont {Freedman}\ and\ \citenamefont
  {McLerran}(1977{\natexlab{c}})}]{Freedman:1976ub}%
  \BibitemOpen
  \bibfield  {author} {\bibinfo {author} {\bibfnamefont {B.~A.}\ \bibnamefont
  {Freedman}}\ and\ \bibinfo {author} {\bibfnamefont {L.~D.}\ \bibnamefont
  {McLerran}},\ }\href {https://doi.org/10.1103/PhysRevD.16.1169} {\bibfield
  {journal} {\bibinfo  {journal} {Phys. Rev. D}\ }\textbf {\bibinfo {volume}
  {16}},\ \bibinfo {pages} {1169} (\bibinfo {year}
  {1977}{\natexlab{c}})}\BibitemShut {NoStop}%
\bibitem [{\citenamefont {Baluni}(1978)}]{Baluni:1977ms}%
  \BibitemOpen
  \bibfield  {author} {\bibinfo {author} {\bibfnamefont {V.}~\bibnamefont
  {Baluni}},\ }\href {https://doi.org/10.1103/PhysRevD.17.2092} {\bibfield
  {journal} {\bibinfo  {journal} {Phys. Rev. D}\ }\textbf {\bibinfo {volume}
  {17}},\ \bibinfo {pages} {2092} (\bibinfo {year} {1978})}\BibitemShut
  {NoStop}%
\bibitem [{\citenamefont {Fraga}\ \emph {et~al.}(2001)\citenamefont {Fraga},
  \citenamefont {Pisarski},\ and\ \citenamefont
  {Schaffner-Bielich}}]{Fraga:2001id}%
  \BibitemOpen
  \bibfield  {author} {\bibinfo {author} {\bibfnamefont {E.~S.}\ \bibnamefont
  {Fraga}}, \bibinfo {author} {\bibfnamefont {R.~D.}\ \bibnamefont
  {Pisarski}},\ and\ \bibinfo {author} {\bibfnamefont {J.}~\bibnamefont
  {Schaffner-Bielich}},\ }\href {https://doi.org/10.1103/PhysRevD.63.121702}
  {\bibfield  {journal} {\bibinfo  {journal} {Phys. Rev. D}\ }\textbf {\bibinfo
  {volume} {63}},\ \bibinfo {pages} {121702} (\bibinfo {year} {2001})},\
  \Eprint {https://arxiv.org/abs/hep-ph/0101143} {arXiv:hep-ph/0101143}
  \BibitemShut {NoStop}%
\bibitem [{\citenamefont {Kurkela}\ \emph {et~al.}(2010)\citenamefont
  {Kurkela}, \citenamefont {Romatschke},\ and\ \citenamefont
  {Vuorinen}}]{Kurkela:2009gj}%
  \BibitemOpen
  \bibfield  {author} {\bibinfo {author} {\bibfnamefont {A.}~\bibnamefont
  {Kurkela}}, \bibinfo {author} {\bibfnamefont {P.}~\bibnamefont
  {Romatschke}},\ and\ \bibinfo {author} {\bibfnamefont {A.}~\bibnamefont
  {Vuorinen}},\ }\href {https://doi.org/10.1103/PhysRevD.81.105021} {\bibfield
  {journal} {\bibinfo  {journal} {Phys. Rev. D}\ }\textbf {\bibinfo {volume}
  {81}},\ \bibinfo {pages} {105021} (\bibinfo {year} {2010})},\ \Eprint
  {https://arxiv.org/abs/0912.1856} {arXiv:0912.1856 [hep-ph]} \BibitemShut
  {NoStop}%
\bibitem [{\citenamefont {Gorda}\ \emph {et~al.}(2023)\citenamefont {Gorda},
  \citenamefont {Paatelainen}, \citenamefont {S\"appi},\ and\ \citenamefont
  {Sepp\"anen}}]{Gorda:2023mkk}%
  \BibitemOpen
  \bibfield  {author} {\bibinfo {author} {\bibfnamefont {T.}~\bibnamefont
  {Gorda}}, \bibinfo {author} {\bibfnamefont {R.}~\bibnamefont {Paatelainen}},
  \bibinfo {author} {\bibfnamefont {S.}~\bibnamefont {S\"appi}},\ and\ \bibinfo
  {author} {\bibfnamefont {K.}~\bibnamefont {Sepp\"anen}},\ }\href@noop {} {\
  (\bibinfo {year} {2023})},\ \Eprint {https://arxiv.org/abs/2307.08734}
  {arXiv:2307.08734 [hep-ph]} \BibitemShut {NoStop}%
\bibitem [{\citenamefont {Fritzsch}\ \emph {et~al.}(2012)\citenamefont
  {Fritzsch}, \citenamefont {Knechtli}, \citenamefont {Leder}, \citenamefont
  {Marinkovic}, \citenamefont {Schaefer}, \citenamefont {Sommer},\ and\
  \citenamefont {Virotta}}]{Fritzsch:2012wq}%
  \BibitemOpen
  \bibfield  {author} {\bibinfo {author} {\bibfnamefont {P.}~\bibnamefont
  {Fritzsch}}, \bibinfo {author} {\bibfnamefont {F.}~\bibnamefont {Knechtli}},
  \bibinfo {author} {\bibfnamefont {B.}~\bibnamefont {Leder}}, \bibinfo
  {author} {\bibfnamefont {M.}~\bibnamefont {Marinkovic}}, \bibinfo {author}
  {\bibfnamefont {S.}~\bibnamefont {Schaefer}}, \bibinfo {author}
  {\bibfnamefont {R.}~\bibnamefont {Sommer}},\ and\ \bibinfo {author}
  {\bibfnamefont {F.}~\bibnamefont {Virotta}},\ }\href
  {https://doi.org/10.1016/j.nuclphysb.2012.07.026} {\bibfield  {journal}
  {\bibinfo  {journal} {Nucl. Phys. B}\ }\textbf {\bibinfo {volume} {865}},\
  \bibinfo {pages} {397} (\bibinfo {year} {2012})},\ \Eprint
  {https://arxiv.org/abs/1205.5380} {arXiv:1205.5380 [hep-lat]} \BibitemShut
  {NoStop}%
\bibitem [{\citenamefont {Aoki}\ \emph {et~al.}(2022)\citenamefont {Aoki} \emph
  {et~al.}}]{FlavourLatticeAveragingGroupFLAG:2021npn}%
  \BibitemOpen
  \bibfield  {author} {\bibinfo {author} {\bibfnamefont {Y.}~\bibnamefont
  {Aoki}} \emph {et~al.} (\bibinfo {collaboration} {Flavour Lattice Averaging
  Group (FLAG)}),\ }\href {https://doi.org/10.1140/epjc/s10052-022-10536-1}
  {\bibfield  {journal} {\bibinfo  {journal} {Eur. Phys. J. C}\ }\textbf
  {\bibinfo {volume} {82}},\ \bibinfo {pages} {869} (\bibinfo {year} {2022})},\
  \Eprint {https://arxiv.org/abs/2111.09849} {arXiv:2111.09849 [hep-lat]}
  \BibitemShut {NoStop}%
\bibitem [{\citenamefont {Annala}\ \emph {et~al.}(2020)\citenamefont {Annala},
  \citenamefont {Gorda}, \citenamefont {Kurkela}, \citenamefont {N\"attil\"a},\
  and\ \citenamefont {Vuorinen}}]{Annala:2019puf}%
  \BibitemOpen
  \bibfield  {author} {\bibinfo {author} {\bibfnamefont {E.}~\bibnamefont
  {Annala}}, \bibinfo {author} {\bibfnamefont {T.}~\bibnamefont {Gorda}},
  \bibinfo {author} {\bibfnamefont {A.}~\bibnamefont {Kurkela}}, \bibinfo
  {author} {\bibfnamefont {J.}~\bibnamefont {N\"attil\"a}},\ and\ \bibinfo
  {author} {\bibfnamefont {A.}~\bibnamefont {Vuorinen}},\ }\href
  {https://doi.org/10.1038/s41567-020-0914-9} {\bibfield  {journal} {\bibinfo
  {journal} {Nature Phys.}\ }\textbf {\bibinfo {volume} {16}},\ \bibinfo
  {pages} {907} (\bibinfo {year} {2020})},\ \Eprint
  {https://arxiv.org/abs/1903.09121} {arXiv:1903.09121 [astro-ph.HE]}
  \BibitemShut {NoStop}%
\end{thebibliography}%

\end{document}